\documentclass[useAMS,usenatbib,epsfig]{mn2e}
\usepackage{graphicx,pdflscape}
\title[Westerlund 2]
{Reddening, Distance, and Stellar Content of the Young Open Cluster Westerlund 2}
\author[Hur et al.]
{Hyeonoh Hur$^{1}$\thanks{E-mail:hhur@sju.ac.kr}, Byeong-Gon Park$^{2}$, Hwankyung Sung$^{1}$\thanks{Corresponding author, E-mail:sungh@sejong.ac.kr}, Michael S. Bessell$^{3}$,  
\newauthor{Beomdu Lim$^{1,2}$\thanks{Korea Research Council of Science and Technology (KRCF) Fellow}, Moo-Young Chun$^{2}$, Sangmo Tony Sohn$^{4}$} \\
$^{1}$Department of Astronomy and Space Science, Sejong University, 209 Neungdong-ro, Gwangjin-gu, Seoul 143-747, Korea\\
$^{2}$Korea Astronomy and Space Science Institute, 776 Daedeokdae-ro, Yuseong-gu, Daejeon 305-348, Korea\\
$^{3}$Research School of Astronomy and Astrophysics, The Australian National University, Cotter Road, Weston Creek ACT 2611, Australia\\
$^{4}$Space Telescope Science Institute, 3700 San Martin Drive, Baltimore, MD 21218}

\begin{document}
\maketitle
\label{firstpage}

\begin{abstract}
We present deep $UBVI_C$ photometric data of the young open cluster Westerlund 2.
An abnormal reddening law of $R_{V,cl}=4.14\pm0.08$ was found for the highly reddened early-type members ($E(B-V)\gid 1.45$),
whereas a fairly normal reddening law of $R_{V,fg}=3.33\pm0.03$ was confirmed for the foreground early-type stars ($E(B-V)_{fg}<1.05$).
The distance modulus was determined from zero-age main-sequence (ZAMS) fitting to the reddening-corrected colour-magnitude diagram of the early-type members 
to be $V_0-M_V=13.9\pm0.14$ (random error) $_{-0.1}^{+0.4}$ (the upper limit of systematic error) mag
($d = 6.0 \pm 0.4 _{-0.3}^{+1.2}$ kpc).
To obtain the initial mass function, pre-main-sequence (PMS) stars were selected by identifying the optical counterparts of
{\it Chandra} X-ray sources and mid-infrared emission stars from the {\it Spitzer} GLIMPSE source catalog.
The initial mass function shows a shallow slope of $\Gamma=-1.1 \pm 0.1$ down to $\log m = 0.7$.
The total mass of Westerlund 2 is estimated to be at least $7,400 M_{\sun}$. 
The age of Westerlund 2 from the main-sequence turn-on and PMS stars is estimated to be $\la 1.5$ Myr.
We confirmed the existence of a clump of PMS stars located $\sim1$ arcmin north of the core of Westerlund 2,
but we could not find any clear evidence for an age difference between the core and the northern clump.
\end{abstract}

\begin{table*}
 \begin{minipage}{160mm}
\caption{The transformation coefficients used to transform the CTIO 1m data to the SAAO standard system}
  \begin{tabular}{c|ccclcccccc}
  \hline
  \hline
Filter & $k_1$      & $k_2$       & $\eta$    & \qquad $\;$Colour index       & $\alpha$ (UT$<5^h$) & $\alpha$ (UT$\gid5^h$) & $\gamma_1$ & $\gamma_2$ & $\zeta$    \\
  \hline
       &            &             & $-0.021 $ & $\quad\qquad\, U-B < 0.7$    &                     &                        &            &            &            \\
$U$    & 0.530      & 0.029       & $-0.200 $ & $\;\;0.7 \lid U-B < 0.9$     & 0.003               & -0.012                 &  0.000     & 0.0043     & 21.869     \\
       & $\pm0.004$ & $\pm 0.007$ & $-0.050 $ & $\;\;0.9\lid U-B\qquad\,\;$  &                     &                        &            &            & $\pm0.011$ \\
  \hline
       &            &             & $-0.3322$ & $\quad\qquad\, B-V < 0.2$    &                     &                        &            &            &            \\
$B$    & 0.283      & 0.027       & $-0.0647$ & $\;\;0.2 \lid B-V < 0.6$     & 0.001               & -0.006                 &  0.012     & 0.0061     & 23.057     \\
       & $\pm0.004$ & $\pm 0.004$ & $-0.2202$ & $\;\; 0.6\lid B-V\qquad\,\;$ &                     &                        &            &            & $\pm0.009$ \\
  \hline
       &            &             &  0        & $\qquad\quad\, B-V < 0.3$    &                     &                        &            &            &            \\
       &            &             & $0.032$   & $\;\;0.3 \lid B-V < 1.43$    &                     &                        &            &            &            \\
$V$    & 0.152      & 0           &  0        & $1.41\lid B-V\qquad\,\;$     & 0                   & $\Delta V^a = -0.016$  &  0.012     & 0.0061     & 23.123     \\
\cline{4-5}
       & $\pm0.003$ &             &  0        & $\qquad\quad\, V-I < 0.3$    &                     &                        &            &            & $\pm0.007$ \\
       &            &             & $0.032$   & $\;\;0.3 \lid V-I < 1.45$    &                     &                        &            &            &            \\
       &            &             &  0        & $1.45\lid V-I\qquad\,\;$     &                     &                        &            &            &            \\
  \hline
$I$    & 0.068      & 0           & $0.038$   &     $\qquad\quad V-I$        & 0                   & -0.004                 &  0.011     & 0.0070     & 22.214     \\
       & $\pm0.003$ &             &           &                              &                     &                        &            &            & $\pm0.007$ \\
 \hline
 \hline
\label{ctio1m_eq}
 \end{tabular}
$^a$Photometric zero-point shift only without time dependence. \\
\end{minipage}
\end{table*}
 
\section{Introduction}
Massive stars are generally expected to form in active star forming regions or young open clusters \citep{po10}.
Observations of young open clusters and OB associations provide important parameters,
such as distance, age, and the initial mass function (IMF), for understanding the formation and evolution of massive stars.
Cores of starburst-type young open clusters are of particular interest since these crowded regions are the most likely sites of
massive star formation through dynamical interactions \citep{mo11,fz12,ok12}.
Hosts of massive stars typically coincide with heavily reddened regions making observations at optical wavelengths difficult.
Nevertheless, if the stars are bright enough to be visible at the shorter optical wavelength ($< 5,000$\AA),
one can adopt the classical techniques for determining the reddening through the use of
e.g. the ($U-B$, $B-V$) two-colour diagram (TCD).
There are only a few young starburst-type open clusters that are suitable for observation from the ground at
the shorter optical wavelengths with a reasonable depth in exposure, and Westerlund 2 (Wd 2) is among them.

Wd 2 was discovered by \citet{we61} as the core of the surrounding H\textsc{ii} region RCW 49.
The cluster is known to be one of the most massive galactic young open clusters
with a total mass of $\sim10^4 M_{\sun}$ \citep{as07}.
Previous photometric or spectroscopic studies have mainly focused on the nature of the massive stars in Wd 2.
A total of 28 O-type stars, two B-type stars, and a massive eclipsing binary system \citep{bo04,ra04} WR 20a (O3 If*/WN6 + O3 If*/WN6 -- \citealt{cw11}) 
had been confirmed in Wd 2. Another massive star, WR 20b (WN6ha -- \citealt{ra11}), is located $\sim$4 arcmin south-east of Wd 2, while
interestingly, \citet{ro11} suggested that three other WR stars, WR 20aa, WR 20c and WR 21a, 
that are located even further from the cluster, may have been ejected from Wd 2 based on their spatial alignment.
\citet{ra07,ra11} presented spectral classifications for 18 O-type stars and two B-type stars in Wd 2.
\citet{va13} added ten O-type stars and four early-type star candidates. 
These spectroscopic studies revealed that Wd 2 contains many massive stars and could be classified as a starburst-type young open cluster.
The nature of PMS stars in Wd 2, however, has not been investigated at optical wavelengths because of their faintness.
However, \citet{as07} presented a deep $JHK_S$ photometric study of Wd 2
and detected PMS stars at these near-infrared (NIR) wavelengths and
\citet{ts07} and \citet{na08} investigated the X-ray properties of early-type and PMS stars in Wd 2 and the RCW 49 region.

Although reddening and distance are the basic parameters needed to understand a cluster,
there are large discrepancies in the distance of Wd 2 among previous photometric studies.
Despite using similar methods -- distance modulus determination via reddening correction,
previous photometric distance measurements of Wd 2 range from 2.8 to 8 kpc,
placing the cluster somewhere along the Sagittarius-Carina arm,
and it is worth noting that even the two most recent studies show a large discrepancy:
4.16 $\pm$ 0.33 kpc by \citet{va13} and 2.85 $\pm$ 0.43 kpc by \citet{ca13}.
This discrepancy is most likely due to the different reddening laws used, the difference in photometry,
and/or isochrones used in the different studies.

In this paper, we aim to determine the basic parameters of Wd 2, reddening law, distance, age, and the IMF, and compare them to the previous determinations.
Based on these parameters and deep optical photometric data, we discuss the stellar population and star formation history of Wd 2.
In section 2, we describe our observations and the data sets used in the analysis. 
In section 3, the reddening law and distance are determined and discussed. 
In section 4, membership selection criteria, the Hertzsprung-Russell (H-R) diagram, age, and the IMF are determined.
The spatial distribution of the cluster members and the OB association in the RCW 49 nebula are discussed in section 5.

\section{Data \label{data_disc}}
\subsection{Optical Photometry\label{optic}}
Deep $UBVI_C$ observations were obtained on 2009 March 28 and 29 with the 8k$\times$8k Mosaic II CCD camera (0.268 arcsec/pix) 
of the 4m Blanco telescope at Cerro Tololo Inter-American Observatory (CTIO).
We used the SDSS $u$ filter for $U$ band photometry and the Harris $BVI$ filter set for $BVI_C$ photometry.
The exposure times were 7 and 150 s in $I$, 10 and 300 s in $V$, 20 and 300 s in $B$, and 100 and 1200 s in $u$.
The average seeing was $\sim0.9$ arcsec.
The observed images were pre-processed by bias subtraction, flat field correction, and cross-talk correction using the \textsc{mscred} package in the \textsc{iraf}\footnote{
Image Reduction and Analysis Facility is developed and distributed by the National Optical Astronomy Observatories which is operated
by the Association of Universities for Research in Astronomy under operative agreement with the National Science Foundation.}.
As Wd 2 is compact enough to be covered with a single chip of the CCD camera (field of view is $17.9\times9.3$ arcmin$^2$), 
we performed the photometric analysis for chip 6 only.

For the photometry of bright stars and for the standardization of $U-B$,
we performed additional observations with the Y4KCam CCD camera and $UBVI_C$ filter set of the 1m telescope at CTIO on 2011 March 5.
The exposure times were 5 and 180 s in $I$, 5 and 180 s in $V$, 10 and 300 s in $B$, and 30 and 600 s in $U$.
As the field of view of the Y4KCam is $19.6\times19.6$ arcmin$^2$, it covers the whole of field of view of Mosaic II CCD chip 6.
The average seeing was $\sim1.7$ arcsec.

\begin{landscape}
\begin{table}
\tiny{
\caption{The combined photometric data obtained with the CTIO 4m and 1m telescopes.}
\label{phot_data}
  \begin{tabular}{p{0.05in}p{0.35in}p{0.45in}cccccccccccp{0.23in}p{0.10in}p{0.10in}p{0.74in}p{0.78in}}
  \hline
  \hline
ID&R.A(J2000).&Dec.(J2000)&$V$&$I$&$U-B$&$B-V$&$V-I$&$\protect\epsilon V$&$\protect\epsilon I$&$\protect\epsilon(U-B)$&$\protect\epsilon(B-V)$&$\protect\epsilon(V-I)$&N(Obs)&Remark$^a$&MSP$^b$&A07$^c$&2MASS&GLIMPSE \\
  \hline
 2167&10:23:18.71&-57:42:20.3&18.479&16.907& 0.570& 1.225& 1.571& 0.005& 0.005& 0.025& 0.007& 0.007&3 3 2 3 3&     &   &    &                 &G284.1620-00.3358\\
 5784&10:24:02.18&-57:45:31.3&16.476&14.367& 0.320& 1.452& 2.109& 0.006& 0.025& 0.014& 0.011& 0.026&2 1 1 1 1&    E&   &3040&                 &                 \\
 5785&10:24:02.20&-57:45:34.3&15.507&13.666&      & 1.550& 1.841& 0.057& 0.046&      & 0.140& 0.073&2 2 0 1 2&   U &   &3041&                 &                 \\
 5790&10:24:02.23&-57:45:02.7&13.892&13.102& 0.201& 0.634& 0.790& 0.002& 0.006& 0.005& 0.004& 0.006&2 2 1 1 2& X   & 91&    &                 &                 \\
 5794&10:24:02.25&-57:45:23.0&17.803&15.244&      & 1.746& 2.559& 0.013& 0.012&      & 0.025& 0.018&3 2 0 2 2&D 2  &   &3065&                 &G284.2707-00.3268\\
 5795&10:24:02.25&-57:45:25.3&16.898&14.741& 0.216& 1.487& 2.172& 0.009& 0.012& 0.096& 0.093& 0.015&2 2 2 2 2& X UE&156&3057&                 &                 \\
 5801&10:24:02.28&-57:45:52.0&19.608&16.277&      & 2.338& 3.332& 0.007& 0.005&      & 0.017& 0.009&3 3 0 1 3& X2  &   &3070&10240227-5745518 &G284.2751-00.3336\\
 5804&10:24:02.29&-57:45:35.4&13.578&11.463& 0.120& 1.395& 2.115& 0.020& 0.082& 0.037& 0.036& 0.084&2 2 1 1 2&DX  E&203&3082&10240230-5745351 &G284.2726-00.3296\\
 5805&10:24:02.29&-57:46:30.9&20.465&17.009&      & 2.428& 3.456& 0.014& 0.012&      & 0.034& 0.018&3 3 0 1 3&DX   &   &3073&10240228-5746309 &G284.2809-00.3427\\
 5806&10:24:02.30&-57:44:31.9&19.075&16.888& 0.602& 1.421& 2.187& 0.011& 0.024& 0.044& 0.016& 0.026&3 3 1 3 3&     &   &    &                 &                 \\
 5807&10:24:02.31&-57:48:54.6&13.280&12.770& 0.284& 0.433& 0.510& 0.004& 0.004& 0.004& 0.006& 0.006&2 2 1 2 2&     &   &    &10240234-5748544 &G284.3022-00.3764\\
 5808&10:24:02.32&-57:44:42.7&21.839&19.727&      & 1.582& 2.113& 0.028& 0.038&      & 0.076& 0.047&1 3 0 1 1&     &   &3083&                 &                 \\
 5809&10:24:02.32&-57:46:47.8&23.614&21.018&      &      & 2.596& 0.126& 0.055&      &      & 0.137&1 1 0 0 1&     &   &3076&                 &                 \\
 5810&10:24:02.32&-57:44:14.8&21.441&19.068&      & 1.846& 2.373& 0.028& 0.015&      & 0.061& 0.032&2 3 0 1 2&     &   &3088&                 &                 \\
 5811&10:24:02.32&-57:45:21.0&18.773&16.668&      & 1.519& 2.120& 0.020& 0.016&      & 0.046& 0.026&2 1 0 1 1& X   &138&3086&                 &                 \\
  \hline
 \end{tabular}
  \begin{tabular}{@{}l@{}}
$^a$ D -- photometric doubles, X -- X-ray emission stars, U -- red leak affected by a close neighbour, E --the early-type members with $E(B-V)\protect\gid1.45$,
e --the foreground early-type stars with $E(B-V)<1.45$, \\ 1 -- class 0/I, 2 -- class II, 3 -- class III   \\
$^b$ ID in the Moffat et al. (1991) \\
$^c$ ID in the Ascenso et al. (2007)\\
This table is available in its entirety in the online journal.
Only a portion is shown here for guidance regarding its form and content.
 \end{tabular}
}
\end{table}
\end{landscape}

\noindent The bias subtraction and flat field correction of these images were applied using the \textsc{ccdred} package in the \textsc{iraf}.
As shutter shading patterns were seen in the shorter exposure images,
we applied the shutter correction to those images with an exposure time shorter than 30 sec\footnote{www.astronomy.ohio-state.edu/Y4KCam/detector.html}.

Instrumental magnitudes of each star on each image were measured by point spread function (PSF) fitting photometry using the \textsc{daophot} package in \textsc{iraf}.
For the CTIO 4m data, an aperture correction was applied to produce the equivalent magnitude for a 5 arcsec radius.
We transformed our $BVI_C$ data to the Stetson version of the Landolt standard system \citep{st00}.
More than 10 images for the standard regions (SA98, SA101, SA104, SA107, SA110, and SA111) 
were obtained for each filter to derive reliable transformation relations for the eight chips of the Mosaic II CCD Camera.
The transformations were described in detail in \citet{li13}.

\begin{table*}
\caption{The mean differences compared to the previous photometric data (this - others).
Photometric doubles and blended stars were not included in the comparisons.
The numbers in parentheses represent the excluded stars due to deviations of more than 2.5$\sigma$ from the mean.
}
  \begin{tabular}{@{}lrrrr@{}}
  \hline
Author        &\citet{msp91}            &\citet{ra07}                &\citet{ca13}              &\citet{va13} \\
 \hline
$\Delta V    $&$ 0.044\pm0.051$, N=48(4)&$-0.016\pm 0.037$, N=230(14)&$ 0.150\pm 0.035$, N=837\enspace ($\enspace$81) &$-0.070\pm0.049$, N=128(17)\\
$\Delta (V-I)$&                         &                          &$-0.041\pm 0.028$, N=785\enspace (116) &$-0.084\pm0.044$, N=129(20)\\
$\Delta (B-V)$&$ 0.066\pm0.046$, N=50(2)&$ 0.011\pm 0.016$, N=230(14)&$-0.002\pm 0.050$, N=740\enspace ($\enspace$76)&$\enspace0.017\pm 0.056$, N=$\enspace$65($\enspace$3)\\
$\Delta (U-B)$&$-0.012\pm0.055$, N=45(2)&                          &$ 0.085\pm 0.049$, N=776\enspace ($\enspace$42)&$\enspace0.144\pm 0.082$, N=$\enspace$59($\enspace$2)\\
 \hline
\end{tabular}
\label{tab_comp_phot}
\end{table*}
For the CTIO 1m data, an aperture correction for a 7 arcsec radius was used and the photometry standardized
by observing many E-region standard stars (E2--E9) from \citet{me89} and
several blue and red standard stars from \citet{ki98} during two observation runs (2010 Oct. 31--Nov. 6 and 2011 Mar. 4 --7).
The adopted transformation relation for the CTIO 1m data was

\begin{equation}
\label{transformation}
M_{\lambda} = m_{\lambda} - (k_{1\lambda} -k_{2\lambda}C)X + \eta_{\lambda}C + \alpha_{\lambda}\hat{UT} + \gamma_{1\lambda}r + \gamma_{2\lambda}r^2 +\zeta_{\lambda} 
\end{equation}
where $M_{\lambda}$, $m_{\lambda0}$, $k_{1\lambda}$, $k_{2\lambda}$, $C$, $X$, 
$\eta_{\lambda}$, $\alpha_{\lambda}$, $\hat{UT}$, $r$, $\gamma_{1\lambda}$, $\gamma_{2\lambda}$ and $\zeta_{\lambda}$ are
the standard magnitude, instrumental magnitude after aperture correction, the primary extinction coefficient, the secondary extinction coefficient, relevant colour index, 
airmass, transformation coefficient, time-variation coefficient, time difference relative to midnight, radial distance from the CCD centre in units of 1,000 pixels,
first order radial variation coefficient, second order radial variation coefficient and photometric zero-point, respectively. 
The adopted relation (equation~\ref{transformation}) is similar to the general form of the transformation relation proposed by \cite{su08,sos0}.
The only difference is the treatment of the spatial variations which was modified to include corrections 
for the radial variations of the CTIO 1m Y4KCAM.
The final adopted coefficients for the CTIO 1m data are summarized in Table~\ref{ctio1m_eq}.
In the $U$ transformation, we found an additional non-linear correction term which depends on the intrinsic colour of the star.
For transformation of the $U$ band of the CTIO 1m data, we adopted
\begin{equation}
\label{U_f}
U = u_0 + f_{[(B-V)_0]} 
\end{equation}
where $U$, $u_0$, and $f_{[(B-V)_0]}$ are the standard $U$ magnitude, the magnitude after the transformation using equation~\ref{transformation}, 
and the non-linear correction term, respectively (See Appendix A for details).

\begin{figure}
\begin{center}
\includegraphics[height=0.30\textwidth]{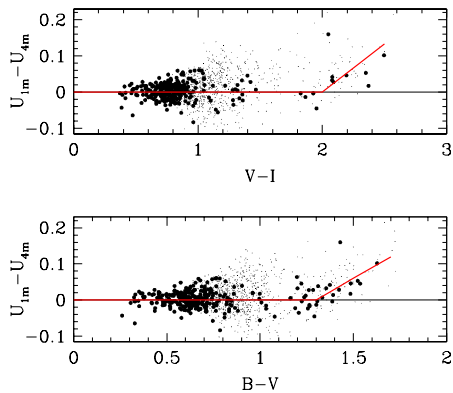}
\caption{The red leak correction relation for $B-V {\protect\lid} 1.7$ and $V-I{\protect\lid} 2.5$ with $\sqrt{\epsilon_{U_{1m}}^2+\epsilon_{U_{4m}}^2}\protect\lid 0.03$.
Large and small dots represent bright stars ($U {\protect\lid} 17.0, B {\protect\lid} 17.5$,
and $V{\protect\lid} 18.0$) and fainter stars, respectively. Photometric doubles were excluded.}
\label{Ured}
\end{center}
\end{figure}

We also found a reddening dependent term in the CTIO 1m $B$ transformation
and a so-called ``red leak" in the SDSS $u$ filter of the CTIO 4m telescope (see also \citealt{li13}). 
The reddening dependent term in the $B$ transformation of the CTIO 1m Y4KCam appears to be caused by the large difference between the CTIO 1m $B$ band 
and the standard $B$ band (see Appendix B for details).
As we used the 4m photometric data for the $B-V$ colour for all stars except saturated stars in the 4m observation and all of these saturated stars 
were less reddened stars, our transformed $B-V$ colours were not affected by the reddening dependent term of the 1m $B$ band. 
But the red leak of the CTIO 4m SDSS $u$ filter resulted in smaller $U$ magnitudes and smaller $U-B$ colours being measured for red stars.
\citet{li13} found that red stars ($B-V>2.0$) were significantly affected by the red leak and their colours were not correctable.
But, although Wd 2 is highly reddened, early-type stars in the cluster are bluer than $B-V=2.0$ ($1.1 \lid B-V \lid 1.7$), 
and were therefore able to be corrected.
We constructed a red leak correction term from comparison of $U$ from the 1m and 4m data (Fig.~\ref{Ured}).
The $U-B$ colour of stars with $B-V > 1.7$ and $V-I > 2.5$ in the CTIO 4m data were discarded from the photometric data 
to avoid problems associated with uncertainty in the derivation of the red leak correction term.

Not only was the $U-B$ colour of a red star affected by the red leak, but close neighbours of a red star were also affected.
Because the red-leak-affected star left an unusual footprint-shaped residual in the PSF-subtracted images, the $U-B$ colour of close neighbours was bluer than its true colour.
We therefore carefully inspected the PSF-subtracted images to identify stars affected in this way. 
The extreme case of such a star was ID 5842 whose $U-B$ was estimated to be smaller by $\sim0.3$ mag.
These stars were excluded in the determination of the reddening law and distance.

Due to a large difference in the seeing between the 4m and 1m observations, we only used the 1m data for the saturated stars in the 4m observations.    
The transformation from CCD coordinates to the equatorial coordinate system was performed by identifying and matching coordinates of
the optical data with the Two Micron All Sky Survey (2MASS) point source catalog \citep{sk06}.
The RMS differences in coordinates after the transformation are 0.15 arcsec in R.A. and 0.17 arcsec in Dec. for the 4m  data 
and 0.14 arcsec in both R.A. and Dec. for the 1m data.
We present our combined photometric data in Table~\ref{phot_data}. 
In the table, the optical counterparts of 2MASS, GLIMPSE, and NIR source from \citet{as07} are also presented.
The information on the photometric doubles, X-ray emission stars, possibly red-leak-affected stars by a bright neighbour, and early-type stars selected in Section~\ref{reddening}
are also included in the table.

\begin{figure}
\begin{center}
\includegraphics[height=0.70\textwidth]{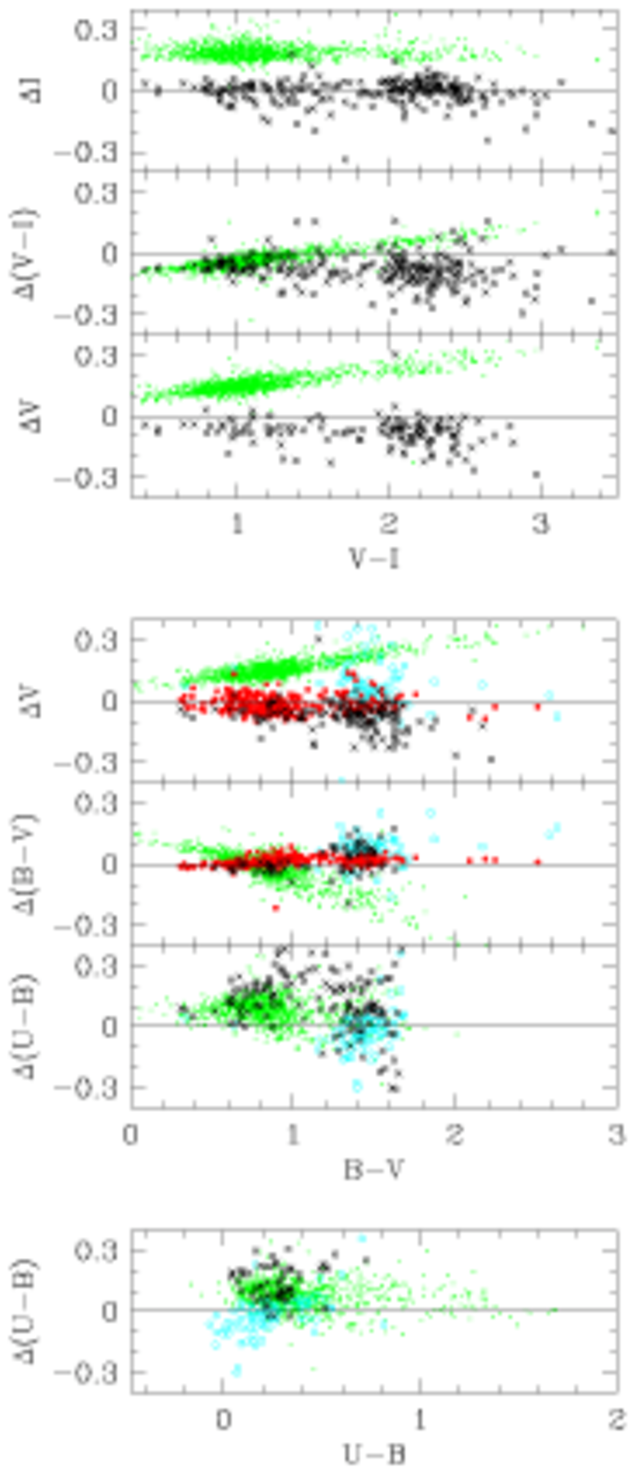}
\caption{
Differences between photometry in the sense (this - others).
Crosses (black)--\citet{va13}, large dots (red)--\citet{ra07}, small dots (green)--\citet{ca13}, and circles (cyan)--\citet{msp91}.}
\label{phot}
\end{center}
\end{figure}

\subsection{Comparison of Photometry\label{com_phot}}
Several photometric data sets have been published for stars in Wd 2 -- $UBV$ photoelectric photometry for 9 stars \citep{mv75};
$UBV$ CCD photometry for 86 stars with the CTIO 0.9m telescope \citep{msp91};
$BV$ CCD photometry: CTIO 1.3m telescope \citep{ra07};
$UBVRI$ CCD photometry with the CTIO 1m telescope and the Swope 1m telescope at Las Campanas \citep{ca13}; and
WFPC2 of the {\it HST} \citep{va13}.
In our analysis of the mean differences between the data sets we excluded photometric doubles in our data, possibly blended stars in \citet{msp91},
stars with photometric errors ($\epsilon \equiv\sqrt{\epsilon_{Ours}^2+\epsilon_{Others}^2}$)$>0.05$ mag,
and any star whose difference deviated by more than 2.5$\sigma$ from the mean difference, to avoid including blended stars or variables.
For comparison with the diffraction limited data of \citet{va13}, 
we selected stars with no blended companions within a radius of 1.0 arcsec and whose positions matched within 0.5 pixel on the Mosaic II CCD camera
(0.135 arcsec) by converting the {\it HST} coordinates to the pixel coordinates on the Mosaic II CCD 
(The RMS differences of the matched stars are 0.16 and 0.12 pixel in R.A. and Dec., respectively). 

Some fainter stars were excluded from the comparison ($V> 18.5$ mag in \citealt{ra07} and \citealt{ca13}; $V>20.0$ mag in \citealt{va13}).
The comparisons between our data and previous photometry are shown in Fig.~\ref{phot} and the mean differences are presented in Table~\ref{tab_comp_phot}.
Our photometry is in good agreement with \citet{ra07}, but the photometry of \citet{msp91} shows large zero-point differences in $V$ and $B-V$.
\citet{ra07} also reported a large difference between their photometry and \citet{msp91}.
The photometry of \citet{ca13} shows the largest differences with a colour dependence in the $V$ magnitude as well as differences in the colours themselves.
Such a systematic difference results in a significantly smaller $E(U-B)/E(B-V)$ ratio than ours (see also Section~\ref{reddening}).
The photometric data of \citet{va13} also show large differences in the $V$ magnitudes and colours. 
As \citet{ra07} described their photometric data as being obtained during the best photometric nights from a multi-night and multi-star observations,
the nice agreement between our data and their data implies that both photometric sets well reproduced the standard system.

\citet{va13} used {\it HST} magnitudes in their quantitative analysis 
but provided $UBVI$ magnitudes transformed to the standard system for comparison with other observers.
They found that the transformed $B$ and $V$ magnitudes were systematically fainter than that of \citet{ra07} 
and concluded that part of the difference was due to crowding. However, we believe that this is not the case. Both our comparison in $V$ based on selected unblended stars, 
and the consistency of the $I$ magnitudes with \citet{va13}, although crowding is much more severe in $I$ than in $V$, rules out crowding as the explanation. 
The most probable reason is a mainly zeropoint difference between the ground-based observations (this work and \citealt{ra07}) and the {\it HST} observations \citep{va13}
due to the difficulty in transforming the {\it HST} magnitudes system to the standard $UBVI$ system as reported in many literatures (e.g. \citealt{sb04,st05,do08}).

\begin{figure}
\includegraphics[height=0.70\textwidth]{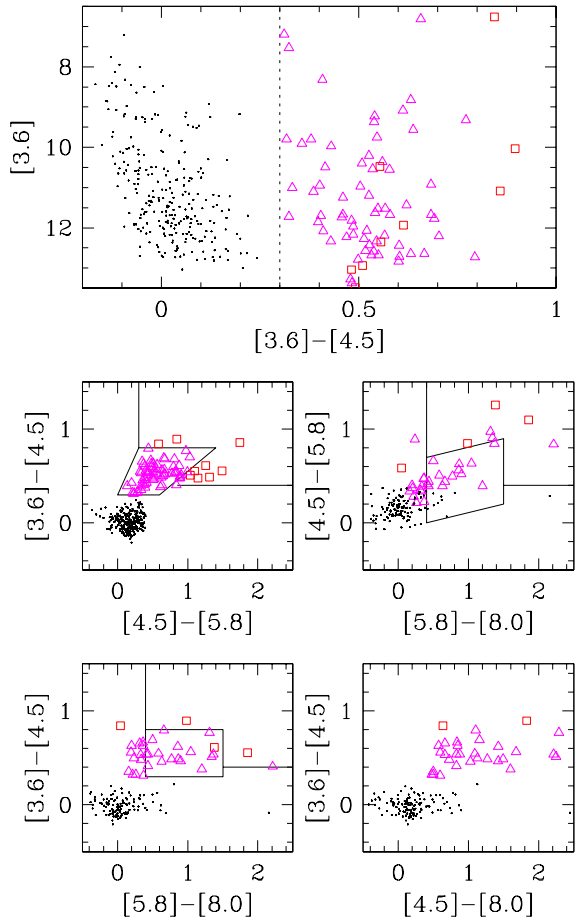}
\caption{The CMD and TCDs of the optically detected GLIMPSE point sources.
Squares (red), triangles (magenta), and dots (black) represent class 0/I, class II, and class III stars, respectively.}
\label{gccd}
\end{figure}

\begin{figure}
\includegraphics[]{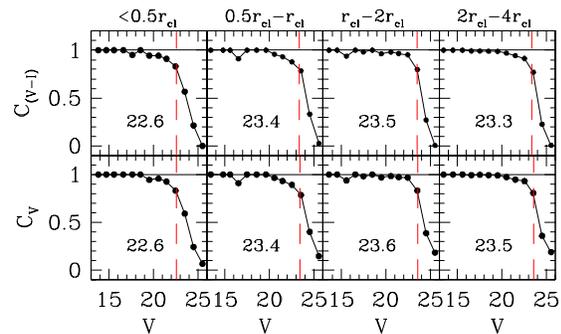}
\caption{The completeness versus $V$ mag.
Dashed lines and numbers in each panel are the $80\%$ completeness limits and its corresponding $V$ magnitudes, respectively.
The upper and lower panels show the completeness in the $V-I$ and $V$ mags, respectively.}
\label{comp_fig}
\end{figure}

\citet{msp91} noted that MSP 203 has a close companion (MSP 444) to the south-east, but only one star ID 5804 was listed in our photometry.
We carefully checked the PSF subtracted image, and 
found a residual to the north-west of ID 5804, but as the star was too close ($\sim0.4$ arcsec) to be decomposed into separated stars,
we measured the object as a single star to avoid large photometric errors in colours of both stars
and added a note `D' (double) to ID 5804 in Table~\ref{phot_data}.
The existence of a close companion to the north-west of the brighter component can be confirmed from the {\it HST} WFPC2 image (see Fig. 7 of \citealt{va13}).
The identity of the counterparts of MSP 203 and MSP444 have been confused.
\citet{ra11}, \citet{ca13}, and \citet{va13} identified the brighter component as MSP 444 based on the relative position in Fig. 2(b) of \citet{msp91},
while MSP 203 is the brighter component in Table 2 of \citet{msp91}.
We infer from the above facts that there was confusion in the illustration of relative positions of MSP 203 and MSP 444 in Fig. 2(b) of \citet{msp91}.
Therefore we assigned the brighter component (the south-eastern component), ID 5804, to be identical to MSP 203.

\subsection{X-ray, near-IR and Spitzer mid-IR data\label{xir}}
As strong X-ray emission is one of the most important characteristics of PMS stars,
we identified the optical counterparts of the {\it Chandra} X-ray point sources from \citet{na08} (result from \textsc{ciao} \textsc{wavdetect} - without any filtering).
A matching radius of 1.0 arcsec for the central $8'\times 8'$ region, and 1.2 arcsec for the outer region was used for the identification
(see section 2.2 of \citet{na08} for the central region).
We searched for the optical counterparts of X-ray sources, and found that 460 X-ray sources have only one star each within the matching radius.
For 54 cases, two stars were within the matching radius so we selected the closest star to the X-ray source as the counterpart. 
Half of the 54 rejected stars were only detected in the $I$ band and were spread over the whole field of view.
The others were relatively bright and predominantly located near the cluster,
but their distance from the X-ray source was more then twice that of the nearest star.
These 54 rejected stars seem to be physically unrelated to the X-ray sources,
and therefore are disregarded as X-ray emission objects.
In summary 514 stars are identified as X-ray emission stars.
One caveat is that as the X-ray luminosity of stars is not strongly correlated with the optical brightness, 
we cannot completely rule out the possibility that a fraction of the rejected stars are the real counterpart of X-ray sources.

We also identified the optical counterparts of the $JHK_S$ photometric data of \citet{as07}, the 2MASS point source catalog \citep{sk06},
and the GLIMPSE point source catalog of \citet{be03}, using a matching radius of 1.0 arcsec.
A total 901, 2565, and 1408 stars were uniquely identified with an object in the NIR sources of \citet{as07}, 2MASS point sources,
and mid-infrared (MIR) GLIMPSE point sources, respectively.
There are 148 cases with two or more NIR objects in \citet{as07} that matched with a star. 
As the angular resolution of the NIR images of \citet{as07} is similar to that of our optical images, 
we assigned the closest NIR object as the counterpart of the optical sources.
The angular resolution of 2MASS images is, on the other hand, much lower than ours, but as we only searched for 2MASS counterparts of optical sources
in the much sparser outer region of Wd 2, only 3 stars were matched with two 2MASS sources within the matching radius. 
Considering the lower angular resolution of the 2MASS images, such cases may be caused by a spurious detection or by errors in the coordinate of the 2MASS point sources,
so we assigned the nearest 2MASS object to the optical source as the 2MASS counterpart.
There are 86 cases with two or more GLIMPSE objects matches for a star. 
As the angular resolution of {\it Spitzer}/IRAC is slightly lower than that of ours,
we assigned the closest MIR source as the counterpart of the star. 
In summary, we identified the NIR and MIR counterpart of 1049+2568 and 1494 stars, respectively. 
However, a few percent of false identification can be expected because of the large difference in the brightness scale between the optical and MIR.

To clarify the evolutionary stage of the young stellar objects (YSOs),
we adopted the loci of class I and II objects in the MIR TCDs as illustrated in Fig. 5 of \citet{su09}.
As all the MIR excess sources were clearly separated from class III/IV locus at $[3.6]-[4.5]=0.3$,
we used that criterion instead of $[3.6]-[4.5]=0.2$.
As a result, we identified 9 class 0/I stars, 66 class II stars, and 259 class III stars with optical counterparts.
Fig.~\ref{gccd} shows the distribution of YSOs in the MIR colour-magnitude diagram (CMD) and TCDs.

\subsection{Completeness Test\label{complet}}
As Wd 2 is one of the densest open clusters in the Galaxy, 
it is very difficult to measure the brightness of its faint stars from seeing-limited images. 
We can therefore expect many faint sources to be missing from ground-based photometry of Wd 2.
To estimate the completeness of our photometry, we performed a completeness test using artificial $V$ and $I$ images.
For the completeness test, we constructed a model cluster using a Monte Carlo method.
The coordinates on the CCD, instrumental magnitude, and instrumental colour were generated using the observed surface density profiles, luminosity function, 
and the CMD, respectively.
Details of this procedure are well described in \citet{su99} or \citet{sb04}.
The PSF photometry, aperture correction, and transformation to the standard system were performed in the same way as for the observed images.
Fig.~\ref{comp_fig} shows the results of the completeness test.
In the figure, the 80\% completeness limit at the cluster central region ($r \lid 0.5 r_{cl}$) is about $V=22.6$ mag from $V$ and $V-I$, 
equivalent to $M_V=1.25$ and $m_{PMS} = \sim4M_{\sun}$, (the faintest stage of the $4M_{\sun}$ PMS evolutionary track of \citealt{si00}) at $V_0-M_V=13.9$ and $E(B-V)=2.00$.
As discussed in \citet{li13}, this completeness limit could be an upper limit because 
there is no proper way to simulate the spikes and bright wings of saturated stars.

\begin{figure}
\begin{center}
\includegraphics[height=0.40\textwidth]{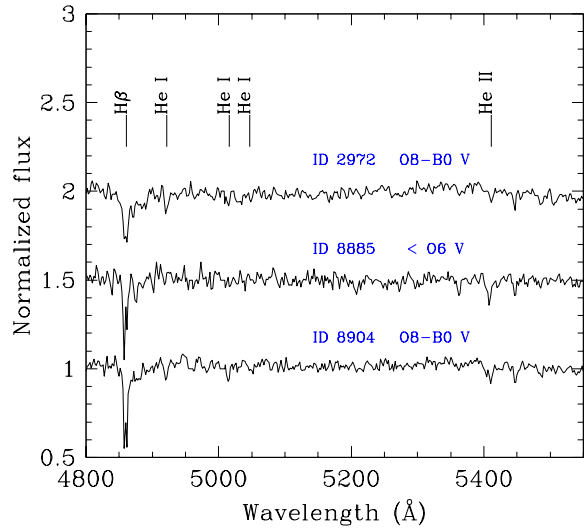}
\caption{The spectral classification of the O-type stars in the RCW 49 nebula.}
\label{Oobs}
\end{center}
\end{figure}

\begin{figure}
\begin{center}
\includegraphics[height=0.40\textwidth]{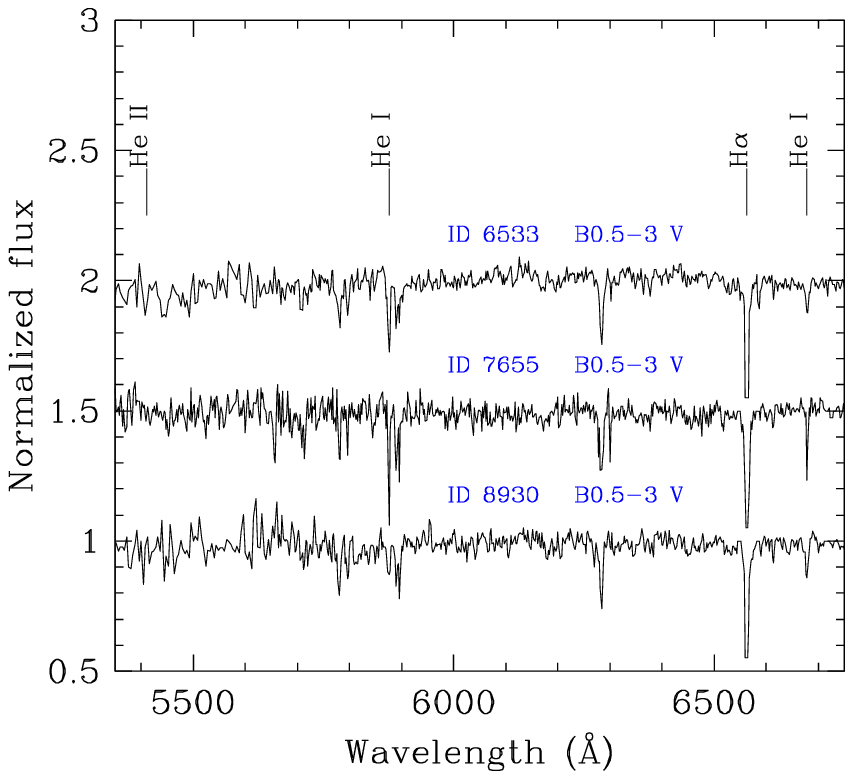}
\caption{The spectral classification of the B-type stars in the RCW49 nebula.}
\label{Bobs}
\end{center}
\end{figure}

\begin{table*}
 \begin{minipage}{160mm}
\caption{Adopted spectral types of the known early-type stars}
\label{sptable}
  \begin{tabular}{@{}llcccccc@{}}
  \hline
ID & Object & $\alpha_{J2000}$ & $\delta_{J2000}$ & RMG \& RSN & CTMB & VKB & Adopted \\
 \hline
5050 & MSP 165         & 10:23:55.18 & -57:45:26.9 &                   &             & O4                & O4 V     \\
5123 & MSP 182         & 10:23:56.18 & -57:45:30.0 & O4 V-III((f))     & O3 V((f))   &                   & O4 V((f))\\
5402 & MSP 223         & 10:23:59.20 & -57:45:40.5 & O7-8              &             &                   & O7.5 V   \\
5515 & MSP 196         & 10:24:00.20 & -57:45:32.5 &                   &             & O8.5              & O8.5 V   \\
5532 & MSP 229         & 10:24:00.35 & -57:45:42.7 &                   &             & O8                & O8 V     \\
5549 & MSP 151         & 10:24:00.48 & -57:45:24.0 & O6 III            & O5 V((f))   & O4                & O4 V     \\
5552 & MSP 44          & 10:24:00.49 & -57:44:44.5 & B1 V + PMS?       &             & late O -- early B & B1 V     \\
5578 & MSP 157b        & 10:24:00.71 & -57:45:25.4 &                   &             & O8                & O8 V     \\
5594 & MSP 157         & 10:24:00.82 & -57:45:25.8 & O6.5 V            & O6 V((f))   &                   & O6.5 V   \\
5615 & MSP 96          & 10:24:00.98 & -57:45:05.4 & B1: + B1:         &             & late O -- early B & B1 V     \\
5629 & MSP 233         & 10:24:01.07 & -57:45:45.7 &                   &             & O9.5              & O9.5V    \\
5644 & MSP 188         & 10:24:01.19 & -57:45:31.0 & O4 V-III          & O4 III      & O3 + O5.5         & O3 V     \\
5673 & MSP 175         & 10:24:01.39 & -57:45:29.6 & O6 V-III          & O4 V((f))   & O4                & O4 V     \\
5692 & MSP 263         & 10:24:01.52 & -57:45:57.0 & O6 V              & O3 V((f))   &                   & O6 V    \\
5743 & MSP 167b        & 10:24:01.88 & -57:45:27.9 &                   &             & O8                & O8 V     \\
5745 & MSP 219         & 10:24:01.89 & -57:45:40.0 &                   &             & O9.5              & O9.5V    \\
5763 & MSP 167         & 10:24:02.06 & -57:45:27.9 & earlier than O6   & O4 III(f)   &                   & O4.5 V   \\
5784 & RSN D           & 10:24:02.18 & -57:45:31.3 & O9.5:             &             &                   & O9.5 V   \\
5804 & MSP 203$^a$     & 10:24:02.29 & -57:45:35.4 & O4-5 V            & O4.5 V      & O4.5              & O4.5 V   \\
5818 & MSP 183         & 10:24:02.36 & -57:45:30.6 & O3 V((f))         & O2 V((f))   &                   & O3 V((f))\\
5823 & MSP 235         & 10:24:02.41 & -57:45:47.1 &                   &             & O9.5              & O9.5V    \\
5827 & MSP 18          & 10:24:02.44 & -57:44:36.1 & O5-5.5 V-III((f)) & O3 III((f)) &                   & O4 V((f))\\
5839 & MSP 183c (RSN C)& 10:24:02.51 & -57:45:31.4 & O7 V              &             & O8.5              & O8.5 V   \\
5842 & MSP 183b (RSN B)& 10:24:02.54 & -57:45:30.4 & O6-7              &             & O8.5              & O8.5 V   \\
5849 & RSN E           & 10:24:02.60 & -57:45:32.2 & O6-7:             &             &                   & O6.5 V   \\
5860 & MSP 199         & 10:24:02.65 & -57:45:34.3 & O3-4 V            & O3 V((f))   &                   & O3.5 V   \\
5874 & RSN A           & 10:24:02.79 & -57:45:30.0 & O8 V              &             &                   & O8 V     \\
6009 & MSP 32          & 10:24:03.79 & -57:44:39.8 &                   &             & O9.5              & O9.5V    \\
6119 & MSP 171         & 10:24:04.90 & -57:45:28.4 & O4-5 V            & O3 V((f))   &                   & O4.5 V   \\
 \hline
\end{tabular}
MSP--\citet{msp91}, RMG--\citet{ra07}, RSN--\citet{ra11}, CTMB--\citet{ca13}, VKB--\citet{va13} \\
$^a$MSP 203 in our data was cross-identified as MSP444 in RMG, CTMB, and VKB (see section~\ref{sp})
\end{minipage}
\end{table*}

\subsection{Spectroscopy and Spectral Classifications\label{sp}}
\citet{ra07,ra11} and \citet{va13} provided the spectral classifications of 27 O-type and two B-type stars;
\citet{ca13} partly revised the spectral types and luminosity classifications of the O-type stars based on the spectra in \citet{ra07}.
We adopted the classification of \citet{va13}, and partly revised the classification of \citet{ra07,ra11}
for our work based on the spectra in \citet{ra07,ra11}.
The star ID 5763 (MSP 167) was originally classified as earlier than O6 by \citet{ra11}.
A weak but obvious presence of He\,{\sc i} $\lambda$4471 implies that the star is O4-O5 V \citep{wf90}, and we adopt the spectral type of the star as O4.5 V.
\citet{ra07} classified MSP 18 (ID 5827) as O5 V--III((f)) and \citet{ra11} presented a spectrum of MSP 18 with better S/N at longer wavelengths
without updating the spectral classification.
Fig.5 of \citet{ra11} shows the absence of He\,{\sc i} $\lambda$4922 ($<$ O5, \citealt{kbm99})
and the presence of obvious He\,{\sc i} $\lambda$5876 ($>$ O3, \citealt{wa80}). We adopt O4 ((f)) for this star.
The luminosity classification criterion of \citet{wf90} for O-type stars is the strength of N\,{\sc iii} $\lambda$4634-4640-4642
relative to that of He\,{\sc ii} $\lambda$4686.
Based on this criterion, we adopted a dwarf luminosity class for all OB stars.
We summarize the adopted spectral types of these stars in Table~\ref{sptable}.

For the spectral classification of the early-type stars outside the cluster radius $r_{cl}$ (see Fig.~\ref{redmap} and Section~\ref{radius}), 
we performed spectroscopic observations for six early-type stars using the WiFeS spectrograph (R=3000) on the 2.3m telescope at Siding Spring Observatory. 
The exposure time was 600 seconds for each star.
As the signal-to-noise ratio between 4000\AA--5000\AA $ $ was too low, 
we used the spectral type classification criteria between 4800\AA--5500\AA $ $ of \citet{kbm99} for O-type stars.
In Fig.~\ref{Oobs}, the stars ID 2972 and ID 8904 showed both  
He\,{\sc i} $\lambda$4922 and He\,{\sc ii} $\lambda$5411 absorption lines, but He\,{\sc i} $\lambda$4922 seems slightly stronger or at least similar to He\,{\sc ii} $\lambda$5411.
Therefore we classified these stars as O8-B0 V.
The star ID 8885 showed no or very weak He\,{\sc i} $\lambda$4922, and He\,{\sc ii} $\lambda$5411 is stronger than in the other two stars.
We conclude that this star is earlier than O6 V which agrees with the SED fitting of \citet{po08} (O5 V at d=4.2 kpc and O3 V at d=6.0 kpc).
Fig.~\ref{Bobs} shows the spectra of the three other early-type stars, ID 6534, ID 7655, and ID 8930. 
The absence of He\,{\sc ii} $\lambda$5411 implies these stars are later than B0. 
The strong He\,{\sc i} $\lambda$5876 and $\lambda$6678 absorptions indicate that these stars are early-B type stars (see also Figure 9 of \citealt{mo13}).
The spectral type classifications are summarized in Table~\ref{sp_obs}. 

\section{Photometric Analysis \label{result}}
\subsection{Cluster Centre and Radius\label{radius}}
To determine the centre of Wd 2, we calculated the coordinates of the highest surface density within a 100 pixel ($\sim0.45$ arcmin) radius.
The centre determined from the $I$ band image was located 22 arcsec north of the centre determined from the $V$ band image.
Although the centre from the $V$ band was located near the core of Wd 2,
we decided to adopt the centre from the $I$ band as the cluster centre 
because many faint stars were detected near the star MSP 18 (ID 5827 = G284.2641-00.3155) in the long exposure $I$ band image.
The coordinates of the centre thus determined are $\alpha _{J2000}=10^{h} 24^{m} 01\fs57$ and $\delta _{J2000}=-57\degr 45\arcmin 08\farcs0$.

\begin{figure}
\begin{center}
\includegraphics[height=0.8\textwidth]{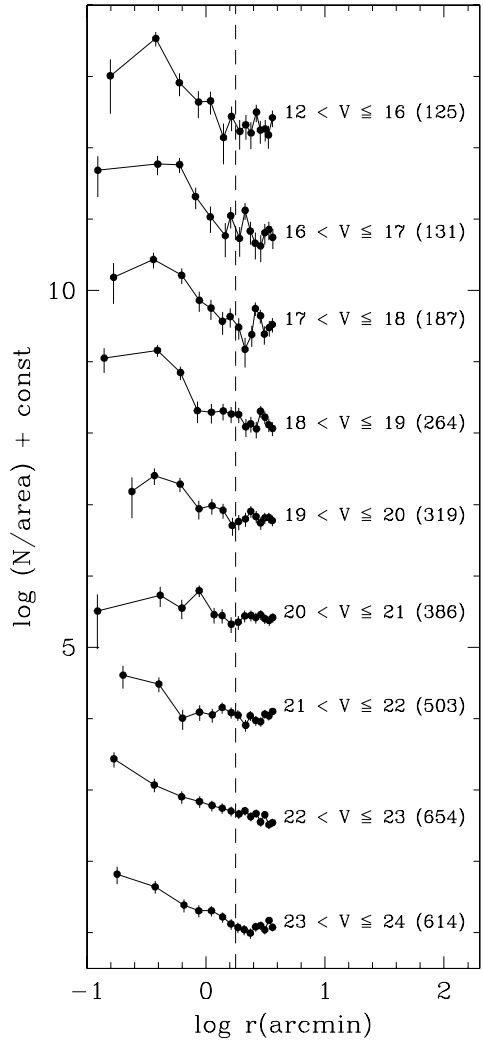}
\caption{The surface density profile of the $V$ magnitude against the radial distance from the cluster centre.
The numbers in parentheses are the numbers of stars in the magnitude bins.
The error bars are based on the Poisson noise.
The dashed line corresponds to $r_{cl}$.}
\label{rad_prf}
\end{center}
\end{figure}

We determined the radius of the cluster ($r_{cl}$) in two ways.
Fig.~\ref{rad_prf} shows the surface density profiles in the $V$ band using the cluster centre determined above.
From the figure, the surface density of the cluster stars approaches that of the field stars at $\log r$(arcmin) $\approx 0.25$.

\begin{figure}
\begin{center}
\includegraphics[height=0.40\textwidth]{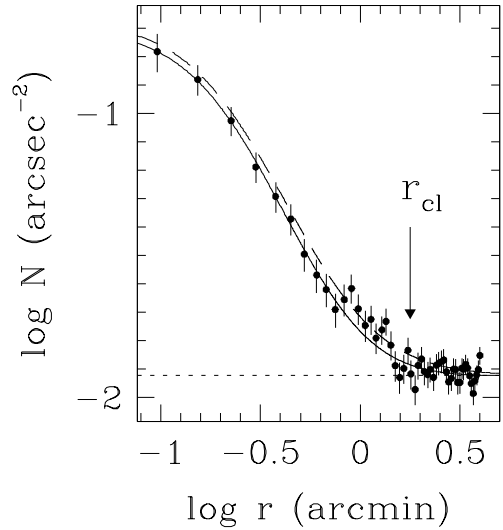}
\caption{Fitting to a King empirical surface density profile for the stars with $V \protect\lid 22.6$ mag.
The solid and dashed lines are the best fit to the King profile and a King profile with a larger tidal radius of $r_t=1000$ arcsec.
The short dashed line represents the background surface density.}
\label{king_fit}
\end{center}
\end{figure}

We also fitted the surface density profile to a King empirical density profile \citep{ki62},
\begin{equation}
\label{king}
f(r)=k[1/\sqrt(1+(r/r_{c})^2)-1/\sqrt(1+(r_t/r_{c})^2)]^2
\end{equation}
where $k$, $r_{c}$, and $r_t$ are the central surface density, core radius, and tidal radius, respectively.
The surface profile fitting was performed for the stars with $V \lid 22.6$ mag. 
The smallest scatter of $\sigma =0.0009$ arcsec$^{-2}$ was found for ($\Delta \alpha,\Delta \delta$) = (0\farcm 00,-0\farcm 42),
$k=0.206$ arcsec$^{-2}$, $r_{c}=12.24$ arcsec, $r_t=290$ arcsec,
and the background surface density of 0.012 arcsec$^{-2}$ (the solid line in Fig.~\ref{king_fit}).
As tidal radius is less meaningful for an open cluster in the galactic disc, there is little difference in the fitting results 
between using the adopted $r_t$ and a larger value ($r_t$=1000 arcsec with the same other parameters).

In addition, it is very difficult to consider Wd 2 as a spherical system because it shows an elongated shape toward the north.
Therefore, we adopt the cluster radius to be $r_{cl}=1.78$ arcmin (equivalent to 3.1 pc at d = 6.0 kpc) for the IMF calculation and the age estimation.
In contrast to $r_{cl}$, the core of the starburst-type cluster Wd 2 seems to be sufficiently concentrated 
and therefore the core radius and central density could be important parameters in characterizing the dynamical status of such clusters.
The core radius of Wd 2 is $r_{c}=0.204$ arcmin (0.36 pc at d=6.0 kpc for m $\ga 3M_{\sun}$), 
which is somewhat larger than NGC 3603 (0.10 pc for m $\ga 5M_{\sun}$, \citealt{sb04}).

\begin{table}
\caption{The Spectral type classifications of the six early-type stars observed with the WiFeS on the SSO 2.3m telescope.
$M_V$ and $(U-B)_0$ were determined after reddening correction.}
\label{sp_obs}
\begin{tabular}{p{0.02\textwidth}p{0.065\textwidth}p{0.079\textwidth}ccl}
\hline
ID   & $\alpha_{J2000}$ & $\delta_{J2000}$ & $M_V$ & $(U-B)_0$ & Sptype   \\
\hline
2972 & 10:23:28.80      & -57:50:04.6      & -5.11 & -1.18    & O8--B0 V  \\
6533 & 10:24:09.43      & -57:43:52.6      & -4.20 & -1.18    & B0.5--3 V \\
7655 & 10:24:23.25      & -57:46:27.6      & -3.61 & -1.19    & B0.5--3 V \\
8885 & 10:24:39.19      & -57:45:21.3      & -5.56 & -1.22    & $<$ O6 V \\
8904 & 10:24:39.48      & -57:45:54.9      & -5.07 & -1.17    & O8--B0 V  \\
8930 & 10:24:39.84      & -57:45:13.1      & -3.58 & -1.10    & B0.5--3 V \\
\hline
\end{tabular}
\end{table}

The presence of many faint sources in the long exposure $I$ image is also seen in the surface density profile, i.e.
a small excess at $\log$ r (arcmin) = -0.1 -- 0.1 is evidently seen in Fig.~\ref{king_fit}.
If the faint sources in the long exposure $I$ image are PMS stars in Wd 2, 
these stars should be also detected and counted in the $K_S$ band.
\citet{as07} also determined the centre of Wd 2 from the starcounts in the $K_S$ band.
There are no other peaks in Fig.7 of \citet{as07}, but the isodensity contour is slightly elongated to the north.
In addition, it is certain that there is a diffuse northern clump around MSP 18 (O4 V((f))) in the $K_S$ image (See Fig.7 of \citealt{as07}).
\citet{va13} also confirmed the northern, secondary concentration of stars from the {\it HST}/WFPC2 images.
The northern clump, in particular, is clearly seen in the X-ray image (see Fig.1 and Fig.2 of \citealt{na08}).
The facts that X-rays are less affected by reddening, and strong X-ray emission is the most important characteristic of young PMS stars, strongly suggest that
these faint sources detected in the northern clump in the $I$ band, are mostly faint PMS stars.
Interestingly, a MIR ring near MSP 18 was found in the {\it Spitzer} MIR images \citep{uz05,va13}. 
While the nature of the MIR ring is still unknown.
MSP 18 is the most likely star related to the ring because the star appears as the most massive star near the ring. 
\citet{va13} suggested that MSP 18 might be a luminous blue variable (LBV).
However, the spectrum of MSP 18 shows no emission lines except for some suspected nebular emission (Fig. 3 of \citealt{ra07} and Fig. 5 of \citealt{ra11}) 
and therefore does not show the typical spectral signature of an LBV.

For the comparison of the stellar content of the core and the northern clump, 
we adopt the centre and radius of the core region and those of the northern clump to be 
($\Delta \alpha$, $\Delta \delta$) = ($0\farcm0$,-$0\farcm42$) and 0.5 arcmin and 
($\Delta \alpha$, $\Delta \delta$) = ($0\farcm0$,+$0\farcm55$) and 0.45 arcmin, respectively.
The centre and radius of the northern clump were set to include most of the early-type stars and PMS stars detected at optical wavelengths
because it is not easy to determine the centre and radius of the northern clump using the starcount method.

\subsection{Reddening\label{reddening}}
The main reason for the large scatter in the distance determined in previous optical photometric studies is the uncertainty in the reddening law,
because the error in $A_V$ is directly related to the error in $V_0-M_V$.
The total extinction in the $V$ band can be determined using the relation,
\begin{equation}
\label{av_standard}
A_V=R_V\times E(B-V).
\end{equation}
The standard reddening law, $R_V=3.1\pm0.2$, has frequently been confirmed for open clusters, not only in the solar neighborhood \citep{su14},
but also beyond the Sagittarius-Carina arm
(Westerlund 1--\citealt{li13}; the galactic centre--\citealt{ri85}).
The normal reddening law of $R_V=3.1\pm 0.2$ was also assumed in the earlier photometric studies for Wd 2 \citep{msp91,ra07}
and the distance to the cluster determined to be $d\sim 8$ kpc ($V_0-M_V\sim14.5$ mag).
But many young open clusters show a larger $E(V-I)/E(B-V)$ ratio \citep{scb00,sb04,hur12} than the normal value, and 
as discussed by \citet{ra07}, higher $R_V$ values should be tested for in the extremely young open cluster Wd 2. 

\begin{figure}
\begin{center}
\includegraphics[height=0.38\textwidth]{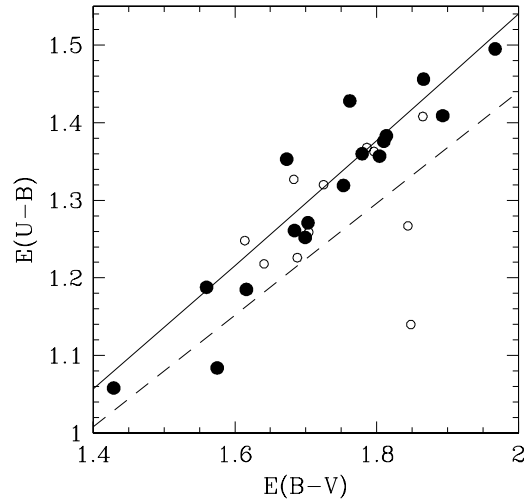}
\caption{$E(U-B)$ versus $E(B-V)$ diagram of stars with O and early-B stars spectral types.
The open circles are probable stars affected by the red leak residual of a close star.
The solid line fitting the data is $E(U-B)/E(B-V) = 0.72 + 0.025 E(B-V)$ and
the dashed line is the normal relation $E(U-B)/E(B-V) = 0.72$.}
\label{eubbv}
\end{center}
\end{figure}

\begin{figure}
\includegraphics[height=0.40\textwidth]{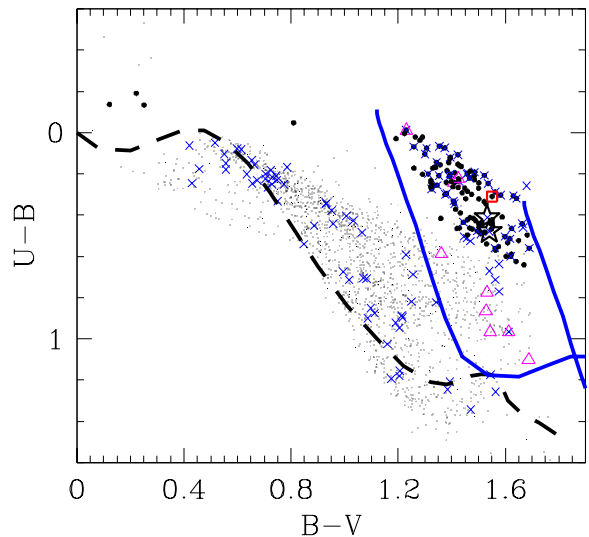}
\begin{center}
\caption{$(U-B)$ versus $(B-V)$ TCD of the stars with
$\epsilon (\equiv\sqrt{{\epsilon (U-B)}^2+{\epsilon (B-V}^2}){\protect\lid} 0.1$ in our field of view.
The solid lines are the reddened ZAMS relations of Wd 2 ($E(B-V)=1.45$ and $E(B-V)=2.00$), and the dashed line is the unreddened ZAMS relation.
The large black dots represent the O and early-B type stars selected in this work, the star symbols represent the Wolf-Rayet stars (WR 20a and WR 20b),
while the crosses, triangles, squares, and small dots represent the X-ray emission stars, class II, class 0/I, and other stars, respectively.}
\label{ubbv}
\end{center}
\end{figure}

Toward the Sagittarius-Carina arm in which Wd 2 is located, NGC 3603 and the young open clusters in the $\eta$ Carina nebula, 
Tr 14 and Tr 16, all show higher $R_V$ values.
Although \citet{sb04} applied a single slope $R_V$ in the colour excess ratio of NGC 3603, 
there seems a slight change in the slope for $E(B-V)\ga 1.3$ in their Fig. 5.
The young stars in Tr 14 and Tr 16 also show a clear difference in the reddening law between the foreground medium and the intracluster medium.
Recently, \citet{ca13} and \citet{va13} reported somewhat larger values of $R_V= 3.85\pm 0.07$ and $R_V=3.77\pm 0.09$, respectively, 
with the assumption of a single $R_V$ value toward the cluster.
But as shown in \citet{hur12}, the $R_V$ value is different for the intracluster region ($R_{V,cl}$) and the foreground ($R_{V,fg}$)
if the dust size distribution in the interstellar medium is different to the size distribution in the cluster, along the same line of sight \citep{ma90}. 
Therefore, the $A_V$ of individual stars should be calculated using the relation, 
\begin{equation}
\label{av_abnormal}
A_V= R_{V,fg}\times E(B-V)_{fg} + R_{V,cl}\times E(B-V)_{cl}
\end{equation}
where $E(B-V)_{fg}$ and $E(B-V)_{cl}$ are foreground reddening and intracluster reddening, respectively.
The early-type stars in Wd 2 show a wide spread in $E(B-V)$ and therefore their $A_V$ and $V_0-M_V$ determination are more sensitive to the adopted $R_V$ value,
particularly for the least and most reddened members.

Using $UBV$ photometry, the reddening $E(B-V)$ of early-type stars are determined from the $(U-B,B-V)$ TCD using the $E(U-B)/E(B-V)$ ratio. 
Firstly, we tested the $E(U-B)/E(B-V)$ ratio toward Wd 2 using the OB stars with known spectral types. 
The colour excesses $E(U-B)$ and $E(B-V)$ were calculated from the difference between the observed colour 
and the intrinsic colour from the spectral type -- colour relation of \citealt{sos0}.
In Fig.~\ref{eubbv}, $E(U-B)/E(B-V)$ toward Wd 2 ($E(U-B)/E(B-V)=0.72+0.025E(B-V)$, solid line) is obviously greater than 0.72 (dashed line)
and therefore we adopt $E(U-B)/E(B-V)=0.72+0.025E(B-V)$.
The investigation of $E(U-B)/E(B-V)$ was described in detail in \citet{sos0}.
\citet{ca13} also determined the $E(U-B)/E(B-V)$ ratio toward Wd 2 using the same method, 
but obtained a different relation of $E(U-B)/E(B-V)=0.63+0.02E(B-V)$. 
The large difference in $E(U-B)/E(B-V)$ between \citet{ca13} and our work seems to be due to large systematic differences between the photometric data.

Fig.~\ref{ubbv} shows the $(U-B,B-V)$ TCD. 
In the TCD, the dashed line is the unreddened ZAMS relation (Table 3 of \citet{sos0}) 
and the (blue) solid lines are the ZAMS relations reddened by $E(B-V)=1.45$ and $E(B-V)=2.00$.
We selected those stars with Johnson's $Q=(U-B) -0.72(B-V) < -0.55$  
and $V \lid 20.5$ mag as early-type members, and marked them as filled circles in the TCD.
Most of the early-type stars show a large spread along the reddening vector between $E(B-V)=1.45$ and $E(B-V)=2.00$,
which implies a large amount of differential reddening across the cluster.
\begin{figure*}
\includegraphics[]{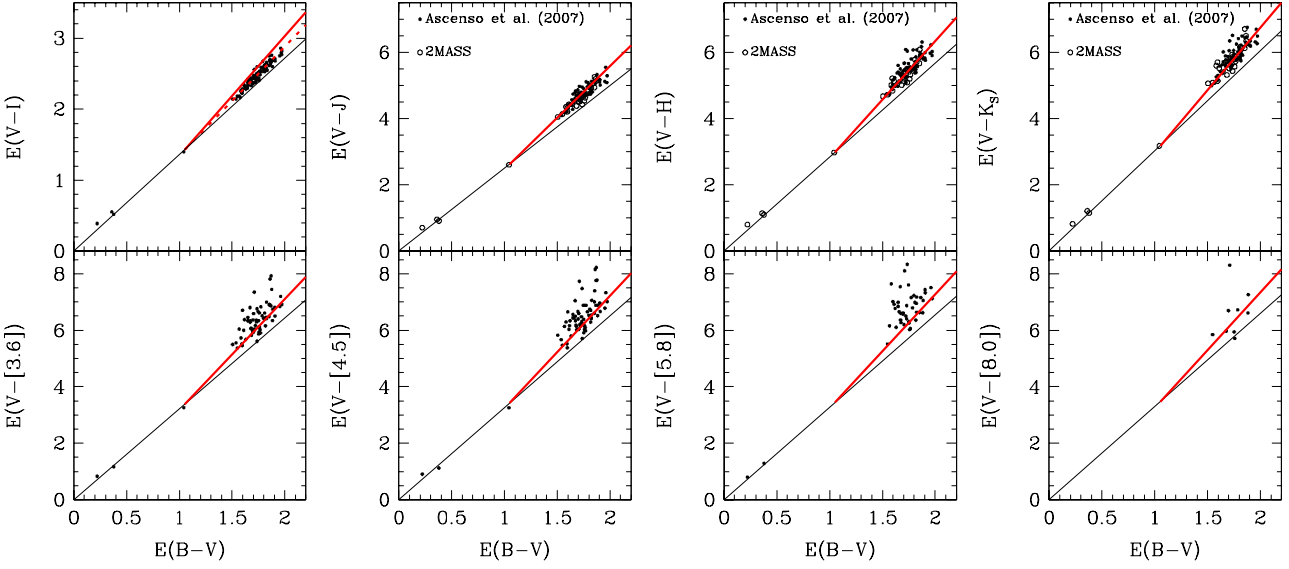}
\caption{The reddening law of Wd 2.
The thin and thick lines are the $E(V-\lambda)/E(B-V)$ ratios for $R_{V,fg}=3.33$ and $R_{V,cl}=4.14$ ($E(B-V){\protect\gid} 1.05$), respectively.
The dashed line corresponds $E(V-I)/E(B-V)=1.53$.}
\label{rv}
\end{figure*}

For the determination of the $R_V$ value, we used the relation between $R_V$ and $E(V-\lambda)/E(B-V)$ ratio for the $I$ and $JHK_S$ of \citet{gu89}.
Fig.~\ref{rv} shows the $E(V-\lambda)/E(B-V)$ relations from the optical to the MIR {\it Spitzer} IRAC bands.
In the figure, there are three less reddened stars and a moderately reddened foreground star.
The less reddened stars may not be suitable to determine the foreground reddening law
because $R_V$ from less reddened stars can be significantly affected by a small error in photometry or a small amount of excess emission.
The moderately reddened early-type star, ID 7081 ($Q=-0.63$ and $E(B-V)=1.04$), shows $R_V=3.33\pm0.03$ from $V-I$, $V-J$, $V-H$, and $V-K_S$  
with a small scatter in both optical and 2MASS data.
Although a direct least-square fit to the colour excess ratios of the stars with $E(B-V)\gid 1.45$ 
in the upper panels of Fig.~\ref{rv} gives a smaller $E(B-V)_{fg}$, 
the amount of foreground reddening should be larger than that of the most reddened foreground early-type star ID 7081.
We adopt the $E(B-V)_{fg}$ to be 1.05.
We also adopt the foreground reddening law to be $R_{V,fg}=3.33$, which agrees with the $R_V$ variation within 3kpc from the Sun \citep{su14}.

The $R_{V}$'s of the intracluster medium of Wd 2 calculated from a least-square fit are $4.07\pm 0.20$, $4.16\pm 0.24$, and $4.23\pm 0.25$
from $V-J$, $V-H$, and $V-K_S$, respectively. 
We adopt the intracluster reddening law, $R_{V,cl}=4.14\pm0.08$ from the weighted mean of the three $R_{V}$'s.
The bottom panels show the $E(V-\lambda)/E(B-V)$ relations for the {\it Spitzer}/IRAC bands calculated from equations (1)--(4) in \citet{su13}.
A large scatter and somewhat larger mean excess ratios above thick lines are seen in all IRAC bands.
The large scatter is due to larger photometric errors in the MIR bands, 
while the large mean colour excess ratios may be related to the excess emission from residual material around these stars.
As the $R_{V,cl}$ from the NIR bands follows the lower edge of the scattered points in the bottom panels,
the colour excess in the IRAC bands also agree with $R_{V,cl}=4.14$.

On the other hand, $E(V-I)/E(B-V)$ indicates a slightly smaller value of $R_{V,cl}=3.76\pm0.10$ (dashed line).
The small $R_{V,cl}$ from $V-I$ is not easily understood because all other colours show a good consistency.
\citet{su13} determined the $R_V$ of the young open cluster NGC 6231 and obtained a consistent result from the optical $I$ band to the MIR IRAC bands.
We carefully cross-checked our results to confirm the coefficients of the CTIO 4m $V-I$ observations (the colour term, spatial variation terms, and photometric zero-point) 
using images for the $\eta$ Carina nebula region (observed in 2008 May using the same standard stars, the same colour terms and spatial variation terms).
From comparison with the photometric data of \citet{hur12} for the stars in the $\eta$ Carina region,  
we confirmed a very good agreement in the $V-I$ colour [$\Delta(V-I)_{4m-Hur}=-0.012\pm0.033$ (N=413)] up to $V-I=3.2$ 
which covers the colour of the most highly reddened early-type stars in Wd 2 ($V-I\sim2.5$). 
Therefore, we conclude the inconsistency of $R_{V,cl}$ is not caused by any problem in the transformation to the standard system.
A possible cause of this discrepancy may be related to the different behaviour of $V-I$ under highly reddened conditions because 
the reddening law in $V-I$ has not been fully investigated for such highly reddened stars.
We adopt $R_{V,fg}=3.33$, $E(B-V)_{fg}=1.05$, $R_{V,cl}=4.14$, and $[E(V-I)/E(B-V)]_{cl}=1.53$ for the reddening correction of cluster members.

Although we believe that the reddening laws of the general interstellar medium and that of the intracluster medium of extremely young open clusters should be 
dealt with separately because the size distribution of dust could be different in the different environments, 
we also calculated a single-value $R_V$ of Wd 2 to compare with previous work.
The single-value $R_V$'s from $V-J$, $V-H$, and $V-K_S$ are $3.62\pm0.19$, $3.65\pm0.22$, and $3.68\pm0.24$ and an average value is $R_V=3.64\pm0.03$ 
which is in agreement with $R_V$ determined by \citet{va13}. 
We discuss the effect of the single-value $R_V$ on the distance modulus determination in the next section.

\subsection{Distance \label{dist}}
As mentioned in the introduction, the distances to Wd 2 derived by previous investigators are very discrepant.
We obtained the distance modulus of Wd 2 from the ZAMS fitting method.
The ZAMS fitting was done by fitting the lower ridge stars in the range of $(U-B)_0$=-0.8 -- -1.05
to avoid the evolutionary effect of early-O stars and the effect of unknown binaries.
It was also considered that the ZAMS relation becomes steeper for early-O stars and therefore more sensitive to a small photometric error in the ZAMS fitting.

\begin{table*}
\tiny{
\caption{$V-NIR$ colours and the reddening-corrected magnitude and colours of the early-type stars derived using equation~\ref{av_abnormal}.}
\label{tab_int}
\begin{tabular}{p{0.15in}p{0.30in}p{0.30in}p{0.30in}p{0.36in}p{0.36in}p{0.36in}p{0.15in}p{0.36in}p{0.36in}p{0.36in}p{0.36in}p{0.37in}p{0.37in}p{0.15in}}
\hline
\hline
 ID   & $V-J$&$V-H$&$V-Ks$&$\epsilon(V-J)$&$\epsilon(V-H)$&$\epsilon(V-K)$&${V_0}^a$ &$(U-B)_0$&$(B-V)_0$&$(V-I)_0$&$(V-J)_0$&$(V-H)_0$&$(V-K)_0$&remark$^{b}$\\
\hline
 5515& 3.853& 4.363& 4.703& 0.035& 0.022& 0.034& 9.772&-1.189&-0.330&-0.360&-0.789&-0.909&-0.999& U \\
 5516& 4.124& 4.564& 4.934& 0.039& 0.050& 0.055&10.047&-1.107&-0.312&-0.335&-0.742&-0.862&-0.939&   \\
 5520& 4.134& 4.794& 5.044& 0.018& 0.036& 0.033&12.471&-0.878&-0.247&-0.273&-0.572&-0.671&-0.720&   \\
 5532& 4.160& 4.680& 5.070& 0.033& 0.038& 0.039& 9.097&-1.147&-0.321&-0.350&-0.766&-0.886&-0.970&D  \\
 5548& 4.036& 4.786& 5.126& 0.012& 0.015& 0.018&11.386&-1.047&-0.297&-0.319&-0.702&-0.821&-0.888&   \\
 5549& 3.921& 4.421& 4.721& 0.025& 0.024& 0.037& 8.287&-1.221&-0.324&-0.353&-0.774&-0.894&-0.980&   \\
 5552& 3.663& 4.309& 4.582& 0.028& 0.029& 0.030& 9.770&-1.125&-0.316&-0.342&-0.753&-0.873&-0.954&  B\\
 5565& 3.964& 4.524& 4.804& 0.016& 0.014& 0.015&12.633&-0.799&-0.227&-0.247&-0.516&-0.607&-0.648&   \\
\hline
\end{tabular}
 \begin{tabular}{l}
$^a$ $M_V$ = $V_0$ - 13.9 mag (the distance modulus).  \\
$^b$ D -- photometric doubles, U -- red leak affected by a close neighbour, B -- known binaries, b -- binary candidates from Table. 1 of \citet{ra11}   \\
This table is available in its entirety in the online journal.
Only a portion is shown here for guidance regarding its form and content.\\
 \end{tabular}
}
\end{table*}

\begin{figure}
\begin{center}
\includegraphics[height=0.50\textwidth]{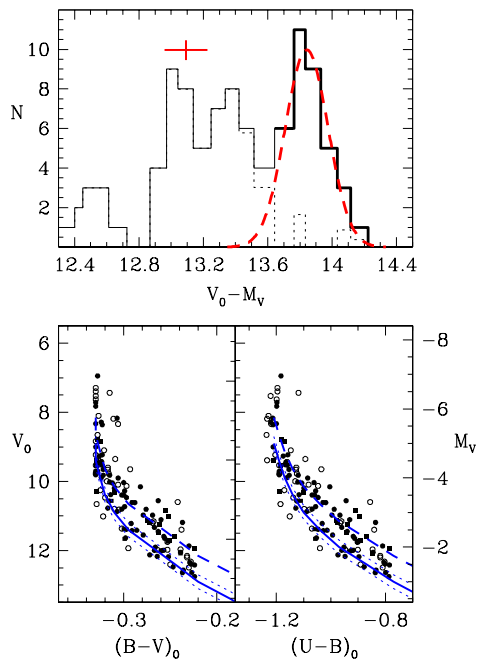}
\caption{The Upper panel shows the distribution of distance modulus of early-type members with $(U-B)_0 \protect\gid -1.05$.
The thick dashed red line corresponds to the Gaussian fit to the distribution centered on $V_0-M_V$=13.84 with $\sigma$=0.13. The red cross
indicates the distance modulus of equal mass binaries ($V_0-M_V=13.84-0.75 \pm 0.13$).
The histogram outlined by the small dashed black lines represents the residual of the original distribution (solid black line) after subtracting the Gaussian fit.
The lower panels show the ZAMS fitted to the early-type members of Wd 2 (in $r_{cl}$, circles) and the RCW49 nebula (outside $r_{cl}$, squares).
Open symbols are photometric doubles, known binaries, stars with $\epsilon (U-B) > 0.03$ mag, 
or stars possibly blended by the PSF-subtracted red leak residual of a close bright star.
The thick solid line, the thick dashed line, and thin dashed lines are ZAMS relations shifted by $V_0-M_V=13.9$ mag, the ZAMS of equal-mass binaries (shifted by $V_0-M_V=13.9-0.75$ mag), and the ZAMS shifted by +0.2 and -0.2 mag from $V_0-M_V=13.9$ mag, respectively.}
\label{zamsfit}
\end{center}
\end{figure}

The bottom panels of Fig.~\ref{zamsfit} show the reddening-corrected CMDs with the ZAMS shifted by the adopted distance modulus.
The solid and dashed lines are the ZAMS relation shifted by the distance modulus of $V_0-M_V=13.9$ mag and
that of the equal-mass binary ($V_0-M_V=13.9 - 0.75$ mag).
There are many stars near the equal-mass binary sequence as well as the single star sequence, which implies a high binary fraction for Wd 2 
(see also the upper panel of Fig.12).

As ZAMS fitting is not a statistical method, it is difficult to estimate the error of the fit.
Because the slope of the ZAMS relation for early-type stars is steep in many CMDs, a small photometric error in colour can significantly affect the ZAMS fitting,
so we used an error-cut of $\epsilon (U-B)\lid 0.03$ mag. 
As the photometric errors of bright early-type stars are small.
We expect the error in distance modulus due to photometric errors to be $\lid0.1$ mag. 
$E(B-V)$ of the stars at the lower ridge in the reddening-corrected CMDs are 1.6 -- 1.88.
The uncertainty in $V_0-M_V$ due to the uncertainty in $R_{V,fg}$ ($3.33\pm0.03$) and $R_{V,cl}$ ($4.14\pm0.08$) is therefore about 0.1 mag.
We conclude that the distance modulus of Wd 2 is 
$V_0-M_V$=13.9 $\pm$0.1 (from photometric error) $\pm$0.1 (from reddening law error) mag ($d = 6.0$ $\pm$0.3 $\pm$0.3 kpc).

In order to fit the ZAMS to a CMD in a statistical way, modeling the expected binarity fraction of the cluster may be useful.
But as other well-observed starburst-type young clusters show complex star formation history \citep{sb04,ro10,sl12,li13} 
and as the CMD morphology of very young open clusters makes it very difficult to discern an age difference of a few Myr,
results from modelling the effects of binaries on a CMD is not very useful without detailed knowledge of the star formation history in the cluster.
In order to estimate the statistical error in the distance modulus determination, 
we present the distance modulus distribution of early-type MS stars in the upper panel of Fig~\ref{zamsfit}.
To avoid the effects of evolution, only stars with $(U-B)_0\gid$-1.05 (mostly B-type stars) were selected.
We also excluded photometric doubles, known binaries, and stars that were blended in the $U$ images. 
A bin size of 0.2 mag was adopted, and a distribution with the same bin size but shifted by 0.1 mag was also constructed
to increase the number of data points. 
Fitting a Gaussian distribution was attempted for points with $V_0-M_V > 13.6$ in order to avoid contamination by unknown binaries. 
The best fitting result was $V_0-M_V=13.84\pm0.13$ mag and the RMS residual after subtraction of the function was 0.45.
That fitting result is quite similar to the distance modulus from ZAMS fitting.
We adopted the distance modulus obtained from ZAMS fitting
to avoid the unknown effect of binarity which can slightly shift the distance modulus from the Gaussian distribution fit.

Spectrophotometric distance is another way to determine the distance of Wd 2.
First, we derived spectrophotometric distances from the stars listed in Table~\ref{sptable} 
using the spectral type--colour and spectral type--$M_V$ relations of \citet{sos0}.
An average spectrophotometric distance modulus, excluding two known binaries (MSP 44 and 96), a possible binary (MSP 18, see \citealt{ra11}), 
and two photometric doubles (MSP 203 and 229), is $13.77 \pm0.66$ mag (24 stars).
However, this method could also be affected by the evolutionary effect of early O-type stars 
and particularly by the presence of unknown binaries if light contribution from secondaries is non-negligible.
An average distance modulus for the stars later than O5 (15 stars) is $14.10 \pm0.36$ mag, which is well consistent with that from the ZAMS fitting.

There are four known eclipsing binaries -- WR 20a, MSP 44, MSP 96, and MSP 223 -- whose light curve were well investigated by \citet{ra07,ra11}.
They derived spectrophotometric distance moduli of these stars to be $14.52\pm0.27$, 14.62, $14.10\pm0.48$, and 14.82 mag (O7.5 V for MSP 223), respectively, 
using the standard reddening law of $R_V=3.1$ and equation~\ref{av_standard}.
As \citet{ra11} calculated spectrophotometric distances of MSP 44, 96, and WR 20a based on the derived parameters from the light curves and the radial velocity curves,
we recalculated the spectrophotometric distance moduli by changing the $A_V$ correction.
The abnormal reddening law of equation~\ref{av_abnormal} gives $\Delta A_V$ (this - \citealt{ra11}) of 
+1.14, +0.69, +0.71, and +0.99 mag for WR 20a, MSP 44, MSP 96, and MSP 223, respectively.
\citet{ra11} did not give the uncertainty in the distance determinations of MSP 44 and MSP 223; however,
if we assume an uncertainty of spectral type to be one sub-class, 
the expected uncertainty is 0.7 mag for MSP 44 (B1 V) and $^{+0.25}_{-0.2}$ mag for MSP 223 (O7.5 V).
The light curves of WR 20a ($13.38\pm0.27$) and MSP 96 ($13.39\pm0.48$) are very similar to that of a contact system, 
and therefore the stars could be brighter than that of a detached system, 
possibly due to the effect of geometric distortion, reflection, and so forth. 
The distance moduli of two detached systems, MSP 44 ($13.93\pm0.70$) and MSP 223 ($13.83^{+0.25}_{-0.2}$), 
are also well consistent with that determined from the ZAMS fitting.

Our distance determination is consistent with the distance determination from the radio emission observations of 
\citet{da07} ($6.0 \pm 1.0 $ kpc), \citet{fu09} ($5.4^{+1.1}_{-1.4}$ kpc), and \citet{be13} ($\sim 6$ kpc), 
although their errors are relatively larger than distances determined in optical techniques, 
while the previous optical photometric studies show a wide spread (3--8 kpc).
The earlier optical photometric studies of \citet{msp91} and \citet{ra07} determined a larger distance of $d\sim8$ kpc.
The cause of this difference in derived distance is definitely due to the difference in the adopted reddening law.
On the other hand, the NIR photometric study of \citet{as07} and the optical photometric study of \citet{ca13} obtained the closer distances of
$2.8$ kpc and $2.85\pm0.43$ kpc, respectively.
Despite the similarity in methods, the main reason of different distances derived by us and by \citet{ca13} is
the different reddening laws used, particularly the different $E(U-B)/E(B-V)$ ratios.

Recently, \citet{va13} determined $R_V$ values and distance moduli of the early-type stars in Wd 2 based on the spectral energy distribution (SED) fitting
using the {\it HST} $UBVI$ magnitudes and the $JHK_S$ magnitudes from \citet{as07}.
They obtained average values of $\left\langle R_V\right\rangle=3.77\pm0.09$, $\left\langle A_V\right\rangle=6.51\pm0.38$ mag,
and $\left\langle d\right\rangle=4.16\pm0.33$ ($\left\langle V_0-M_V\right\rangle=13.10\pm 0.17$) after the binary effect correction.
As they obtained a higher $R_V$ value than $R_V=3.1$, 
the discrepancy in distance determination seems not to be due to the reddening correction.  
We discuss the origin of this discrepancy in the next section.

\subsection{Systematic Errors in the Reddening Correction and Distance} 
We examined how the possible systematic errors in the reddening correction and the ZAMS relation affect the distance determination.
For the test, we performed the same procedure for the de-reddening, $R_{V,fg}$ and $R_{V,cl}$ determination, and $A_V$ calculation using other's empirical ZAMS relations.
The empirical ZAMS relations from \citet{me81} and \citet{sc82}, which include both $(U-B)_0$ versus $(B-V)_0$ and $(U-B)_0$ versus $M_V$ relations, were used in the test.
As \citet{me81} presented $(U-B)_0$ versus $(B-V)_0$ relation down to only $(U-B)_0=-1.05$, 
stars with $(U-B)_0 < -1.05$ were excluded when using the ZAMS from \citet{me81}.

We found only small differences (less than 0.01 mag in average) between the $E(B-V)$ derived from our result and both ZAMS relations.
The reddening laws derived from these ZAMS relations are $R_{V,fg}=3.34\pm0.01$ and $R_{V,cl}=4.12\pm0.08$ from \citet{me81} 
and $R_{V,fg}=3.34\pm0.01$ and $R_{V,cl}=4.13\pm0.08$ from \citet{sc82}, 
and so we confirm that the effect of using different ZAMS relations is very small in determining the reddening law and $A_V$ (less than 0.01 mag in $A_V$).
On the other hand, $M_V$ of \citet{sc82} is systematically brighter than those of \citet{sos0} and \citet{me81}.
For the stars with $(U-B)_0$=-0.8 -- -1.05, the average difference in $M_V$ (\citealt{sos0} - others) is $0.40\pm0.05$ mag for \citet{sc82} and $ 0.00\pm0.04$ mag for \citet{me81},
and this would cause a systematic error in distance modulus determination.
As the ZAMS of \citet{sc82} is also brighter than other empirical ZAMS relations 
(e.g. $\sim$0.32 mag brighter than the ZAMS of \citealt{tu76} for $(B-V)_0=$-0.2 -- -0.3), 
the upper limit of the systematic errors in distance modulus would be smaller than this.  

The lower limit of the systematic error in the empirical ZAMS relation is difficult to estimate
because there are no published fainter empirical ZAMS relations for $(B-V)_0$=-0.2-- -0.3 
and the slope of the ZAMS for  $(B-V)_0<-0.3$ is too steep.
For $(B-V)_0$=-0.2-- -0.3, we found average differences (\citealt{sos0} - others) 
of $0.50 \pm 0.08$ from \citet{sc82}, $-0.01 \pm 0.02$ from \citet{me81},
$0.18 \pm 0.10$ from \citet{tu76}, and $0.51 \pm 0.12$ from \citet{bl63},
all positive differences.
However, the presence of unknown intrinsic uncertainty in the ZAMS relation, 
such as that due to systematic difference in distance or photometric scatter in the sample used in the determination of the ZAMS, is possible.
For these reasons, the ZAMS relation itself could involve an uncertainty of $\sim$0.1 mag, 
and this could contribute another source of the systematic error in our distance determination.
Therefore the distance modulus and its possible error is $V_0-M_V=13.9$ mag $\pm0.1$ (from the photometric error) 
$\pm0.1$ (from the reddening law error)  $_{-0.1}^{+0.4}$ 
(the possible systematic error from the empirical ZAMS) (d=6.0 kpc $\pm0.3$ $\pm0.3$ $_{-0.3}^{+1.2}$).

The application of one reddening law for the foreground ($R_{V,fg}=3.33$) and another
for the intracluster medium ($R_{V,cl}=4.14$), instead of a single reddening law, does affect
the $A_V$ and $V_0-M_V$ derived for individual early-type stars.
If we adopt a single reddening law ($R_V=3.64\pm0.03$ from our data),
the expected difference in distance modulus is -0.13 mag for the least reddened star [$E(B-V)$=1.45],
and +0.15 mag for the most highly reddened star [$E(B-V)$=2.00] and -0.09 mag for the stars at the lower ridge in the reddening-corrected CMDs.
Therefore an additional error of -0.1 mag is expected if the reddening law toward Wd 2 can be represented by a simple single-slope reddening law.

Another source of uncertainty in the distance determination is the difference in the adopted relations between $R_V$ and colour excess ratios.
A recent study of the NIR reddening law of \citet{fm09} presented a modified relation,
\begin{equation}
\label{rv_fm09}
R_V=1.36E(V-K_S)/E(B-V)-0.79
\end{equation}
which gives a higher $R_V$ for $R_V>4.0$ stars.
From this relation and the same ZAMS fitting method, we obtained $R_{V,cl}=4.44\pm0.31$ and $V_0-M_V=13.7$ mag from the two-slope reddening law and
$R_V=3.76\pm0.29$ and $V_0-M_V=13.7$ mag from the single-slope reddening law.
However, we should be cautious in adopting Equation~\ref{rv_fm09}.
\citet{fm09} noted that their work was based on only 14 stars and more NIR data are necessary to verify the relation.
Some of the stars they used to derive the above relation, such as the stars in the $\eta$ Carina nebula,
are members of the region which shows a clear two-slope reddening law \citep{hur12}.
Although \citet{fm09} is a recent study on the $R_V$ determination,
their relation is determined under the assumption of a single $R_V$ which could have a more serious effect on stars with a large $R_V$.
If $R_V$ is correlated with the grain size distribution in the line of sight as they noted,
it is not easy to neglect the effect of different grain sizes between the general interstellar medium and the intracluster medium
because young open clusters are formed in the dense core of a giant molecular cloud
whose grain size distribution differs from that of the general interstellar medium.

The recent distance determination of \citet{va13} is $\sim$0.8 mag smaller than our result.
As we used the empirical ZAMS relation of \citet{sos0} in the reddening determination
while \citet{va13} used the Padova theoretical isochrones,
it is worth checking how the adopted isochrones could affect the derived reddening law.
We used the ZAMS from the Padova isochrones\footnote{http://stev.oapd.inaf.it/cgi-bin/cmd} for Z=0.019 and
the 1 Myr isochrone from the Geneva models\footnote{http://obswww.unige.ch/Recherche/evoldb/index/Isochrone/} (with rotation) for Z=0.014.
The Padova isochrone was transformed to the photometric system of \citet{bs90} and \citet{be88}.

The foreground reddening law for the reddened foreground star ID 7081 was calculated to be $E(B-V)=0.99$ and $R_{V,fg}=3.35\pm0.05$ and
$E(B-V)=1.06$ and $R_{V,fg}=3.29\pm0.07$ from the Padova and Geneva isochrones, respectively.
We adopted $E(B-V)_{fg}$ to be 1.0 and 1.05 for the Padova and Geneva isochrones and obtained the $R_{V,cl}$ to be $4.20\pm0.06$ and $4.07\pm0.09$, respectively.
We also calculated the reddening law of the 1.5 Myr Padova isochrone and obtained a quite similar reddening law of $R_{V,cl}=4.22\pm0.13$.
The $E(B-V)$ values determined for the early-type stars using the Padova isochrone were systematically smaller by $\sim$0.06 mag
which would result in a larger distance modulus by $\sim$0.2 mag,
while the $E(B-V)$ values calculated from the Geneva isochrone were quite similar to ours within 0.01 mag.
Despite a slightly smaller $E(B-V)$ from the Padova isochrone, the derived reddening laws are very similar.
We conclude that the effect due to the application of theoretical isochrones on the reddening law is also small.

Despite the expected difference in distance modulus from the reddening difference ($\sim$+0.2 mag for the Padova isochrone),
ZAMS fitting indicated a smaller distance modulus of 13.5 mag from the Padova isochrone, while 14.2 mag was obtained from the Geneva isochrone.
Therefore one reason for the difference in distance modulus between \citet{va13} and our work seems to be the difference 
between the empirical and theoretical isochrones on the observational CMDs and TCD.

The large difference in the distance moduli between the isochrones seem to be
due to the different colour-temperature relations and bolometric corrections adopted in the isochrones,
because the theoretical isochrones show a large discrepancy
in the observational diagrams (CMDs and TCD, this was also confirmed by \citealt{tw99} for a solar mass star),
while there are no significant differences in the MS band among several published stellar evolutionary models in the H-R diagram \citep{ma13}.
For a given colour, the Geneva isochrone is 0.5--1 mag brighter in the ($M_V,(U-B)_0$) CMD and more than 0.75 mag fainter in the ($M_V,(B-V)_0$) than the Padova isochrone.
The ZAMS of \citet{sos0} is located nearly midway between the two isochrones in the both CMDs and the $(M_V,(V-I)_0)$ CMD.
Additionally, the $(B-V)_0$ colour of the hottest stars of the Geneva isochrone reaches $(B-V)_0=-0.37$, while the bluest $(B-V)_0$ colour of \citet{sos0} is -0.33 mag.
\citet{ek12} noted that they used the colour-temperature relations of \citet{fl77,bo81,sc82},
but the relations gave the bluest $(B-V)_0$ colour as -0.33 mag up to mid-O stars.
If they extrapolated the empirical colour-temperature relations for early-O stars, the intrinsic $B-V$ colour would be excessively blue.
Therefore we also tested the Geneva isochrone with a constraint of $(B-V)_0=-0.33$ for the hottest stars,
but both the reddening law and the distance modulus determined from the Geneva isochrone showed no significant difference
due to the bluest colour of the $(B-V)_0$.

The other reasons of the disagreement in distance with \citet{va13} seems to be the reddening law and binary correction. 
\citet{va13} obtained an average value of $R_V= 3.77 \pm 0.09$ from  SED fittings for individual early-type stars.
The difference in $R_V$ between our adopted values and their value results in 0.1 -- 0.3 mag smaller $A_V$ for $E(B-V)$= 1.5 -- 2.0 than our reddening law.

The binary correction could also be an expected source of the difference in distance modulus 
because we fitted the ZAMS to the lower ridge line of the early-type members, 
while \citet{va13} took the average distance modulus of individual early-type stars and adopted a binary correction of -0.12 mag
based on the estimated light contribution from secondaries of O-type and early-B type stars in Cyg OB 2 \citep{ki12}.
However, their observed binary fraction of B-type stars (8/69) are obviously lower than O-type stars (12/44) 
while \citet{va13} used 27 O-type stars and two B-type eclipsing binaries in their analysis.
This lower binary fraction could be attributed to the lower S/N of spectra of B-type stars due to their faintness 
or the intrinsic difference in binary fraction between O-stars and B-stars.
As there is no well-studied binary fraction or light contribution from secondaries for such a starburst-type young open cluster, especially its core,
the possibility of a higher binary correction than $\sim$-0.1 mag to the distance modulus still cannot be ruled out.
In order to test the correction for the binary effect on the O-type stars in their sample, 
we ran a simple Monte Carlo simulation for the OB-type stars used in the \citet{va13}
using the intrinsic binary fraction of O-type stars (69\%) from \citet{sa12} with a mass ratio ($q$=0--1) for the binary systems.
For two B-type eclipsing binaries, MSP 44 and 96, we adopted V magnitudes for the secondaries based on the second minimum of the light curves \citep{ra07}.
The minimum, maximum and mean values of distance modulus shift from 1000 simulation runs were -0.08, -0.45, and $-0.24\pm0.06$ mag, respectively.
As \citet{sa12} obtained the binary fraction of O-type stars in young open clusters with ages up to $\sim$5 Myr \citep{su13},
we also tested the binary correction using the intrinsic ZAMS binary fraction
(binary fraction of 75\% for the O-type stars and 37\% for the binaries with an O-type secondary), (see supplementary C.1 of \citealt{sa12})
and obtained -0.15, -0.50, and $-0.30\pm0.06$ mag, respectively.
Both cases indicate that a somewhat larger correction cannot be ruled out for Wd 2.
Binary correction is one of the most difficult problems in stellar astronomy and
it is not easy to estimate the binary effect on the distance modulus of the O-type stars in Wd 2.
Therefore, the reasons for the difference in the distance modulus between our result and \citet{va13} could be
due to a combination of the different ZAMS relations used ($\sim$0.4 mag), the difference in $A_V$ calculation($\sim$0.2 mag), 
and possibly to a smaller binary correction ($\sim$0.2 mag).

We note that in Figure 19 of \citet{va13}, about half of their early-type members are fainter than the ZAMS line
which also supports a larger distance modulus being more appropriate for Wd 2.
As the ZAMS relation refers to the colour (equivalently spectral type) -- luminosity relation for ``single" MS stars with age ``zero",
we should fit the lower ridge line of the MS band, and as a result, a larger distance modulus is more appropriate for Wd 2.

\subsection{Colour-Magnitude Diagrams\label{cmd_text}}
\begin{figure*}
\begin{center}
\includegraphics[height=1.22\textwidth]{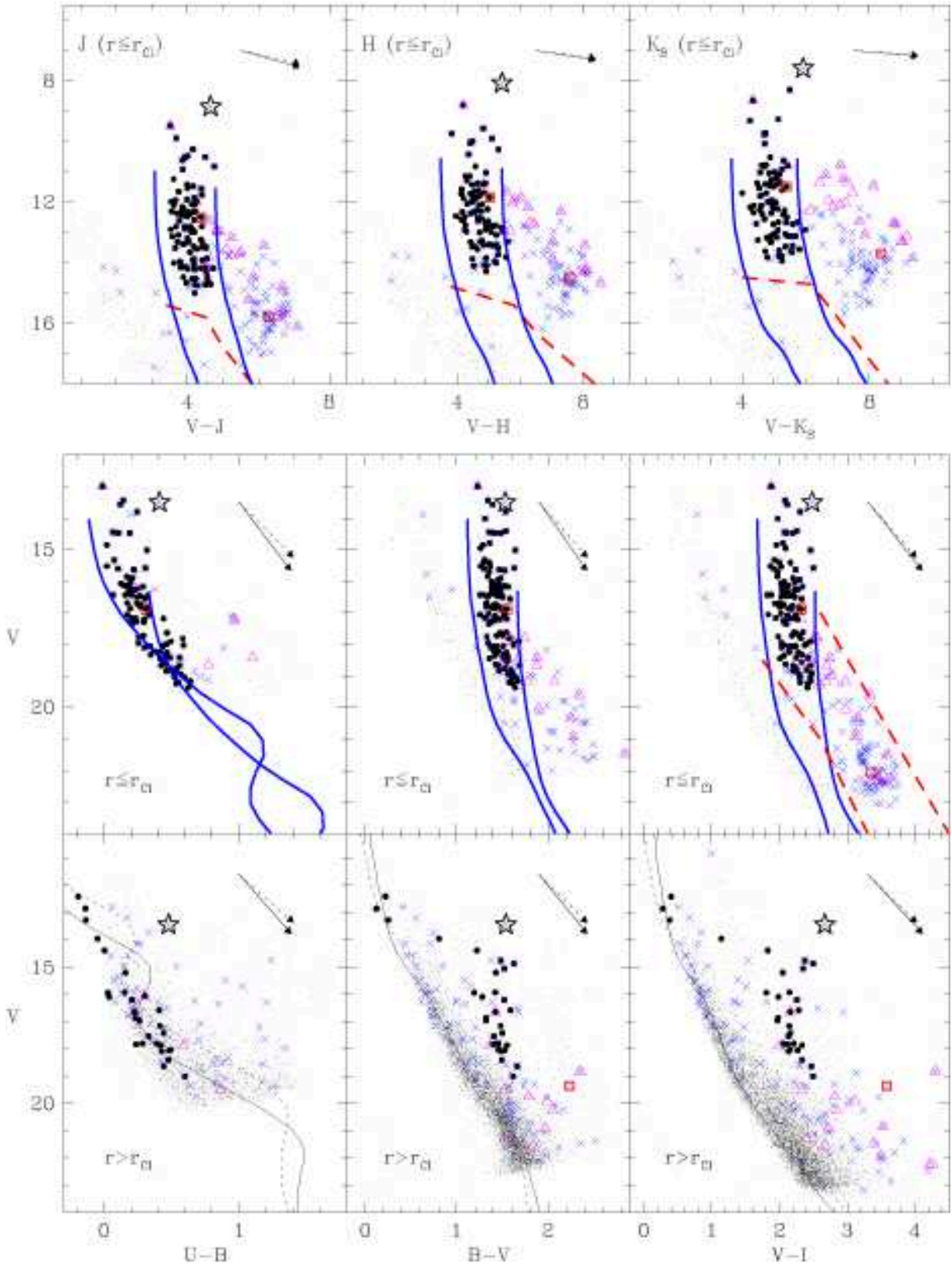}
\caption{Colour-magnitude diagrams of the stars with
$\epsilon (\equiv\sqrt{{\epsilon (mag)}^2+{\epsilon (colour)}^2}\,){\protect\lid} 0.1$.
The thick solid lines in the upper and middle panels are the reddened ZAMS relations of Wd 2 ($E(B-V)=1.45$ and $E(B-V)=2.00$, respectively and $V_0-M_V=13.9$).
The thin solid and dashed lines in the lower panels are the reddened ZAMS relation of $E(B-V)=0.36$ and $V_0-M_V=11.8$, and $E(B-V)=0.20$ and $V_0-M_V=11.0$, respectively.
The thick dashed lines in the upper panels represent the dividing line between field and cluster stars.
The thick dashed lines in the $(V,V-I)$ diagram represent the PMS locus.
The solid and dashed arrows indicate the reddening vectors of intracluster ($R_{V,cl}$) and foreground ($R_{V,fg}$), respectively, for $E(B-V)=0.5$ mag.
Other symbols are same as Fig.~\ref{ubbv}.}
\label{cmd}
\end{center}
\end{figure*}

We present several optical-NIR CMDs in Fig.~\ref{cmd}.
To construct the CMDs in the upper panels, we used the $JHK_S$ data from the NTT observations for the central $4.2\times4.1$ arcmin$^2$ area of Wd 2 \citep{as07} 
and 2MASS point source catalog data \citep{sk06} for the outer region.
The early-type members within $r_{cl}$ show a clearly separated red sequence in the ($V,B-V$) and ($V,V-I$) diagrams.
Most of the early-type stars are located within
the reddened ZAMS relations with $V_0-M_V=13.9$ and $E(B-V)=1.45$ and $2.00$ (thick lines).
There are several early-type stars with $r>r_{cl}$ including WR 20b (WN6ha). 
These early-type stars in the outer region appear to be at the same distance as the early-type stars within $r_{cl}$.
We discuss the spatial distribution of these early-type stars outside $r_{cl}$ in detail in Section~\ref{halo_early}.
Many class 0/I stars, class II stars, and X-ray emission stars show a clear PMS sequence in the ($V,V-I$) diagram.
The dashed lines in the ($V,V-I$) CMD (middle right) represent the PMS locus 
which was set to include most class 0/I stars, class II stars, and X-ray emission stars.
The MS and PMS sequences are more distinctly separated in the combined optical-NIR CMDs than the optical-only CMDs.
We also set the cluster-field boundary in the $(J,V-J)$, $(H,V-H)$, and $(K_S,V-K_S)$ diagrams (the dashed lines in the upper panels) 
to separate the cluster members from field stars.

An early-type star ID 5908 ($V=16.891$, $B-V=1.551$, 2MASS J10240304-5746542, GLIMPSE G284.2858-00.3474) was classified as a class 0/I star.
There are four early-type stars whose YSO class are class II [ID 5181, ID 5827, ID 8566, and ID 8930 (early-B, see Fig.~\ref{Bobs})].
These stars might be early-type stars with a PMS secondary star or Herbig Be stars with circumstellar discs.
In the ($V,V-I$) CMD for $r>r_{cl}$, there are several class 0/I stars, class II stars, or X-ray emission stars 
which seem to be slightly redder than the red boundary of the PMS locus.
Most of these stars appear to be PMS stars in the RCW 49 nebula. 
Four YSO class II stars (ID 5142, ID 5408, ID 5650, and ID 6003) were detected in $U-B$ and are located within the PMS locus.
Three of the four stars (ID 5408, ID 5650, and ID 6003) appear to be NIR excess stars, and
ID 5408 ($K_S=10.80$) and ID 5650 ($K_S=10.64$) are brighter in the $K_S$ band ($K_S\gid12$) than any other PMS stars.
All these four stars are located within $r_{cl}$.
There are several more bright class II objects with similar colours and magnitudes in the ($V,V-I$) ($2.2<V-I<2.8$ and $17.0<V<19.0$)
and ($K_S,V-K_S$) ($6.0<V-K_S<8.0$ and $10.5<K_S<12.5$) diagrams.
These stars seem to be Herbig Be stars with a strong NIR or MIR excess from a circumstellar disc.
A few class II stars outside $r_{cl}$ are located below the lower boundary of the PMS locus and seem likely to be PMS stars with an edge-on disc         
because a PMS star with a thick edge-on disc, such as W90 in NGC 2264 \citep{su97}, can show much fainter $V$ magnitude due to the 
grey extinction caused by large-size dust in the circumstellar disc.
There are a few stars with very red $V-I$ colour (ID 2973, ID 3412, and ID 7360).
These stars seem to be field late-type giants and were excluded from the membership list.
\begin{figure*}
\begin{center}
\includegraphics[height=0.34\textwidth]{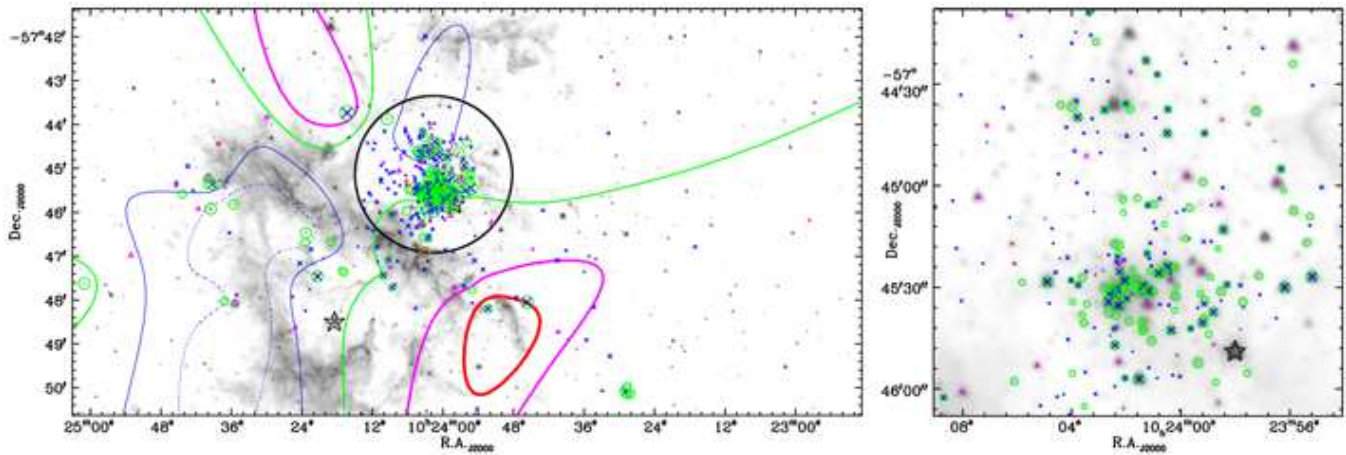}
\caption{Left -- The reddening map and the spatial distribution of early-type stars and PMS members on the reversed {\it Spitzer} 3.6$\mu m$ image.
The dashed blue, solid blue, green, magenta, and red contours represent $E(B-V)=$1.65, 1.70, 1.75, 1.80, and 1.85, respectively.
The black solid circle indicates the radius of Wd 2.
Star symbols and green circles are WR stars and early-type stars, respectively.
Magenta triangles, red squares, blue crosses, magenta crosses, and red crosses represent the PMS members of class II, class 0/I, X-ray emission,
YSO class II with X-ray emission, and YSO class 0/I with X-ray emission, respectively.
The size of the symbols is based on the mass determined from the H-R diagram.
Right --  A close-up map for Wd 2 with $1\times1$ arcmin$^2$.}
\label{redmap}
\end{center}
\end{figure*}

The solid and dashed thin lines in the bottom panels of Fig.~\ref{cmd} are the less reddened ZAMS relations with  
[$E(B-V)= 0.36$, $V_0-M_V=11.8$] (solid lines) and [$E(B-V)= 0.20$, $V_0-M_V=11.0$] (dashed lines) which are appropriate for most foreground field stars. 
As Wd 2 is located toward the Sagittarius--Carina spiral arm, these sequences appear to be the sequences of foreground field stars in the arm.
The reddening $E(B-V)$ and distance modulus of the foreground spiral arm is well consistent 
with the foreground early-type stars of the young open clusters Tr 14 and Tr 16 in the $\eta$ Carina nebula \citep{hur12}.
The three less reddened foreground early-type stars are well fit with [$E(B-V)= 0.36$, $V_0-M_V=11.8$].
A moderately reddened foreground early-type star, ID 7081 ($V=13.96$, $B-V=0.81$, and $E(B-V)=1.04$), 
that is located between the foreground sequence and the cluster sequence, appears to be an early-B field star. 
There are some bright red X-ray emission stars ($15<V<18$, $0.8<U-B<1.4$) which appear to be very close late-type stars ($V_0-M_V\sim8.0$, $d\sim 400 pc$).

\section{H-R Diagram\label{H-R}}
\subsection{Membership Selection and Reddening Correction\label{member}}  
The early-type stars between $E(B-V)=1.45$ and $E(B-V)=2.00$ were selected as the early-type members.
The most massive stars, WR 20a (O3 If*/WN6 + O3 If*/WN6) and WR 20b (WN6ha), were also included in the member list.
Stars with $Q \lid -0.55$, or without a $U-B$ colour but located within the MS locus ($V\lid20.5$ mag and between $E(B-V)=1.45$ and $E(B-V)=2.00$) 
in all CMDs, were selected as MS member candidates.
Stars with X-ray emission, those of class 0/I, or II within the PMS locus (Fig.~\ref{cmd}), were selected as PMS members 
and other stars with no membership criterion within the locus, were selected as PMS member candidates. 
Stars slightly redder than the red boundary of the PMS locus with the membership criteria, were also selected as PMS members.

The reddening $E(B-V)$ of the early-type stars were derived individually using the reddening law determined in section~\ref{reddening} 
and equation~\ref{av_abnormal}.
For the reddening correction of PMS stars, we constructed a reddening map using the spatial distribution of early-type members.
Fig.~\ref{redmap} shows the contours of $E(B-V)$ overlying the composite {\it Spitzer} MIR image.
Within $r_{cl}$, $E(B-V)$ increases from north to south.
In the region outside $r_{cl}$, there are an insufficient number of early-type stars and the contours therefore represent a moderate tendency to a spatial reddening variation.

\subsection{Colour-T$_{eff}$ Relation and Bolometric Correction\label{C-T}}
For the construction of the H-R diagram, we transformed the intrinsic colours, spectral type, and absolute magnitude
to effective temperature (T$_{eff}$) and bolometric magnitude. 
The adopted spectral type--$T_{eff}$ relation, colour--$T_{eff}$ relations, and bolometric corrections are well summarized in \citet{sos0}.
For two WNh stars, we adopted $T_{eff}=43000$ K and $\log L/L_{\sun}=6.28$ (a total luminosity of the binary system from \citealt{ra11})
for WR 20a (O3 If*/WN6 + O3 If*/WN6) and $T_{eff}=45000$ K and $\log L/L_{\sun}=6.18$ for WR 20b (WN6ha) (from Table 2 in \citealt{cr07}).
The spectral type was adopted as the T$_{eff}$ indicator from O3 V to O7 V stars and 
both the spectral type and the $(U-B)_0$ -- $T_{eff}$ relation were adopted, with an equal weight, for late-O stars.
We adopted $(B-V)_0$ and $(U-B)_0$ as $T_{eff}$ indicators for other MS stars and $(V-I)_0$ for PMS stars.
For the stars affected by a red leak residual of a close bright star, the $(U-B)_0$ colour was not used as a $T_{eff}$ indicator to avoid an overestimation of T$_{eff}$.

\begin{figure}
\includegraphics[height=0.40\textwidth]{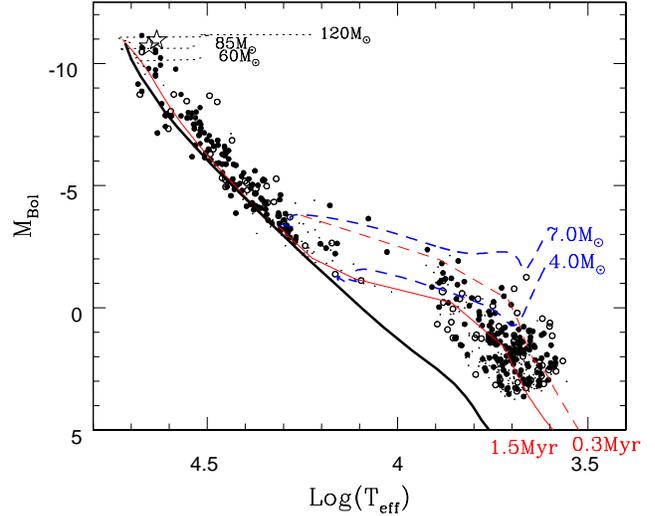}
\caption{The H-R diagram of Wd 2.
The filled circles, open circles, dots, and stars represent the members of Wd 2 ($r\protect\lid r_{cl}$), the members of the RCW 49 nebula ($r>r_{cl}$),
the candidates of Wd 2, and the evolved stars (WR 20a and WR 20b), respectively.
The thick solid and long dashed lines are the ZAMS of \citet{ek12} and the evolutionary tracks of \citet{si00} with the masses to the right, respectively.
The thin solid lines represent the 1.5 Myr isochrones interpolated from the stellar evolutionary tracks of \citet{ek12} ($M_{Bol}>-3$)
and the PMS evolutionary tracks of \citet{si00} ($M_{Bol}\protect\lid-3$), respectively.
The thin long dashed line represents the 0.3 Myr isochrone from the PMS evolutionary tracks.
The thin short dashed lines indicates the stellar evolutionary tracks before the MS turn-off for 60, 85, and 120 $M_{\sun}$.
}
\label{hrd}
\end{figure}

The H-R diagram of Wd 2 ($r\lid r_{cl}$) and the stars in the RCW 49 nebula ($r>r_{cl}$) are shown in Fig.~\ref{hrd}.
The PMS evolutionary tracks of $4M_{\sun}$ and $7M_{\sun}$ from \citet{si00} are drawn for comparison with the positions of PMS stars in the H-R diagram.
The isochrones for 1.5 Myr and 0.3 Myr from the PMS evolutionary tracks and 1.5 Myr from the stellar evolutionary models of \citet{ek12} are also shown in the H-R diagram. 

\subsection{The Initial Mass Function\label{IMF_text}}
The initial mass of individual stars were determined from comparison of their position in the H-R diagram with the theoretical evolutionary tracks.
The stellar evolutionary solar metallicity models with rotation of \citet{ek12} were adopted for the MS stars and 
the PMS evolutionary solar metallicity tracks of \citet{si00} were adopted for the PMS stars.
We assumed the initial masses of WR 20a and WR 20b to be the most massive stars in the stellar evolutionary models ($120M_{\sun}$) 
because the initial mass of these stars is generally very difficult to determine due to rapid evolution and mass loss during MS and post-MS evolution.
For the calculation of the initial mass function (IMF), only the mass of the primary stars were considered 
because although the binarity of several of the early-type stars of Wd 2 had already been studied by \citet{ra11}, 
photometry-based mass determination cannot take into account the mass of a secondary star.

\begin{figure}
\begin{center}
\includegraphics[height=0.7\textwidth]{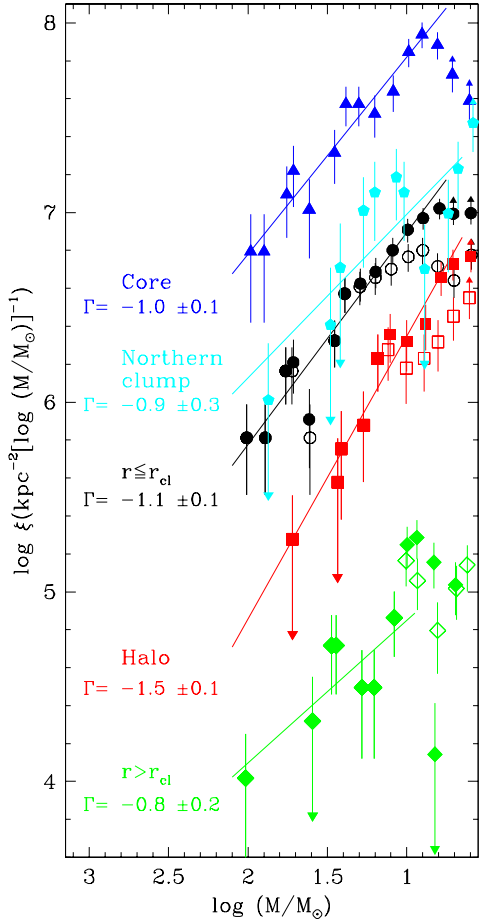}
\caption{The IMF of the core (triangles), the northern clump (pentagons), $r\protect\lid r_{cl}$ (circles), 
the halo (open squares), and $r> r_{cl}$ (diamonds).
The filled and open symbols represent the IMF calculated for the members+candidates with field subtraction 
and members only without the subtraction, respectively.
The error bars are based on $\sqrt N$.}
\label{imf}
\end{center}
\end{figure}

Fig.~\ref{imf} shows the IMFs $\xi$($\equiv N/\Delta \log m $/area) calculated using the members and candidates. 
The IMFs were calculated with a bin size of $\Delta \log m = 0.2$
and the IMFs shifted by 0.1 in $\log m $ were also calculated to minimize the binning effect.
The error bars are based on the Poisson noise of each point.
Only one massive star (ID 5785, $\log m \sim 1.6$ in figure~\ref{imf}) was selected as a massive member candidate due to the lack of $(U-B)$ colour 
because the star was strongly affected by the red leak residual of neighboring bright star (MSP 203+444) in the $U$ images.
The estimated completeness limit is $\log m = 0.6$ ($m = 4 M_{\sun}$),
but the apparent completeness limit of the IMF of the core seems to be $\log m > 0.7$ 
and the completeness of the $\log m $ = 0.6 -- 0.8 bin seems to be significantly lower, 
probably because the completeness limit in the core could be underestimated by the crowding effect of saturated stars in the $V$ and $I$ bands.
Therefore, we calculated the slope of the IMF for $\log m > 0.7$ (down to the $\log m $ = 0.7 -- 0.9 bin) for the core and within $r_{cl}$ and 
$\log m > 0.6$ (down to the $\log m $ = 0.6 -- 0.8 bin) for the northern clump and the halo 
(the area within the $r \lid r_{cl}$ excluding the core and the northern clump, see Section~\ref{halo_pms}). 
The slope of the IMF for stars in the RCW 49 nebula ($r > r_{cl}$) was calculated down to $\log m>0.9$ ($m > 7 M_{\sun}$)
to avoid significant field star contamination in the PMS locus.
The most massive bins of the IMF for $r>r_{cl}$, the northern clump, and the halo were calculated with a bin size of 
$\Delta \log m =$ 0.4, because of the small number of very massive stars.

As open clusters lie in the galactic plane, 
the membership selection and correction for field star contamination are one of the most difficult problems in the IMF calculation.
In the case of Wd 2, most MS members are O- and early B-type stars selected in the $(U-B,B-V)$ TCD,
and therefore we expect the field contamination of early-type MS stars to be negligible.
But the membership selection of PMS stars could be contaminated by field stars in the galactic plane.
Hence the IMF outside $r_{cl}$ would show an excess of low mass stars ($\log m < 1.0$) due to severe contamination by field stars.
In order to correct for field star contamination in the PMS locus, 
a control field which represents the field population toward the cluster is necessary.
However, it is very difficult to find such a control field near very young open clusters 
because the spatial distribution of field stars is not uniform due to the irregular distribution of molecular clouds.
We tried to subtract the number of field interlopers from the CMD of the field region [$\alpha < 10^h23^m24\fs1$ ($\Delta \alpha < -5\farcm0$)] using the areal ratio, 
but obtained a highly over-subtracted IMF due to the different number density of the field population.

We tried another method to estimate the field contribution in the CMD by counting the number of stars in the field star locus 
[stars bluer than the reddened ZAMS by ${E(B-V)=1.45}$ for $V\lid 19$ or stars fainter than the lower boundary of the PMS locus for $19<V\lid 22.6$]
as well as that in the member locus [within two reddened ZAMS lines [($V\lid 19$) or within the PMS locus ($19<V\lid 22.6$)].
A total of 1570 and 370 stars in the field star locus were counted for the field region and $r\lid r_{cl}$, respectively, and 
26 and 219 stars in the member locus with $m \gid 4M_{\sun}$ were counted for each region, respectively. 
The expected field contamination calculated from the field region is therefore only 2.8\%. 
The most massive field interloper was calculated to be 6.8 $M_{\sun}$ if the star is classified as a member candidate. 
Therefore, field star contamination is expected mainly for PMS stars.
The effect of field contamination on the IMF slope is calculated to be negligible ($\sim 0.01$ for $r\lid r_{cl}$). 

Asymptotic-giant-branch (AGB) stars are also known as stars with MIR excess emission. 
If the expected number of AGB contamination is non-negligible, we should statistically subtract them from the MIR emission stars.
However, the expected number of AGB contamination within $r_{cl}$ is estimated to be less than 0.2 stars using equation (10) of \citet{ro08},
and therefore we can neglect the contamination of AGB stars.
The contamination of background non-stellar sources is also expected to be negligible for optically detected stars,
because of the large extinction due to background nebulae.
\citet{ge11} simulated the extragalactic contamination of X-ray sources in the $\eta$ Carina nebula to be fainter than $J=21$ mag.
Wd 2 is of similar age or younger than the clusters in the $\eta$ Carina nebula because 
the evolution stages of the most massive stars (WR 20a and WR 20b) seem to be earlier than $\eta$ Carinae (LBV) \citep{sm08}.
As the two regions are also located at similar galactic longitudes, extragalactic sources are too faint to be detected at optical wavelength, at least within $r_{cl}$.
As a result, we conclude again that field star contamination of the PMS membership selection within $r_{cl}$ to be very small and negligible.

The slopes of the IMF within $r_{cl}$, the core, the northern clump, and the halo
were determined to be $\Gamma=-1.1 \pm 0.1$ (s.e.), $\Gamma=-1.0 \pm 0.1$ (s.e.), $\Gamma=-0.9 \pm 0.3$ (s.e.), 
and $\Gamma=-1.5 \pm 0.2$ (s.e.), respectively.
The massive stars are mainly located in the core, while faint stars are also concentrated in the northern part 
near MSP 18 (O4 V).
The IMF of the core shows the highest surface density.
We conclude that the global IMF of Wd 2 (within $r_{cl}$) is slightly flat ($\Gamma=-1.1\pm0.1$). 

The slope of the IMF of Wd 2 is much shallower than that of other star forming regions such as 
NGC 2264 ($\Gamma=-1.7 \pm 0.1$, \citealt{su10}) or IC 1848 ($\Gamma=-1.6 \pm 0.2$, \citealt{li14a})
and slightly shallower than the $\eta$ Carina nebula region ($\Gamma=-1.3 \pm 0.1$, \citealt{hur12}) or NGC 1893 ($\Gamma=-1.3 \pm 0.1$, \citealt{li14b}). 
It is similar to that of NGC 6231 ($\Gamma=-1.1 \pm 0.1$, \citealt{su13} or the Arches cluster($\Gamma=-1.1 \pm 0.2$, \citealt{es09}), 
but is slightly steeper than other starburst-type young open clusters, 
Westerlund 1 ($\Gamma=-1.0\pm0.1$, \citealt{li13}) or NGC 3603($\Gamma=-0.9\pm0.1$, \citealt{sb04}).

\subsection{Total Mass\label{totalmass}}
We calculated the total mass of Wd 2 within $r_{cl}$ by dividing the mass range into three sections, 
$\log m \gid 0.7$, $0.7 > \log m \gid -0.3$, and $\log m < -0.3$.
For $\log m \gid 0.7$, we counted all members and candidates within $r_{cl}$.
For $0.7 > \log m \gid -0.3$, we extrapolated the slope of the IMF ($\Gamma=-1.1$) down to $\log m = -0.3$.
As there is no well-studied IMF of starburst-type clusters down to very low-mass stars,
we calculated the total mass for $\log m < -0.3$ by adopting two well-studied cases, 
case 1) -- the Kroupa IMF \citep{kr01} and case 2) -- the IMF of NGC 2264 \citep{su10} by shifting $\log \xi +1.0$.
We obtained a similar total mass in the two cases, 7,421 $M_{\sun}$ for case 1) and 7,415 $M_{\sun}$ for case 2).
While a high binary fraction is expected for Wd 2, we could not take into account the mass of the secondary stars 
because the photometric method cannot take into account the binary fraction, as a function of stellar mass and mass ratio distribution, for a given mass range.
Therefore the total mass would be much more massive and the total mass of $\sim 7,400 M_{\sun}$ obtained above could be a lower limit.
\citet{as07} converted the NIR luminosity function into a mass function using the mass--luminosity relation and obtained a similar lower limit to the total mass. 

The total mass of starburst-type young open clusters generally exceed $\sim 10^4 M_{\sun}$.
The R136 cluster in the LMC appears to be the most massive young open cluster ($\sim 10^5 M_{\sun}$, \citealt{an09}) in the local group.
The total mass of starburst-type clusters in the Galaxy seems to be less massive than the R136 cluster 
[The Arches cluster ($\sim 2\times 10^4 M_{\sun}$, \citealt{es09}), Westerlund 1 ($ > 5\times 10^4 M_{\sun}$, \citealt{li13}), 
and NGC 3603 (0.7 -- 1.6$\times 10^4 M_{\sun}$, \citealt{ha08,ro10,st06})].
From the comparison of the total mass with NGC 3603, 
the only well observed galactic starburst-type cluster with both photometric and spectroscopic observations at short wavelength, 
we conclude that the total mass of Wd 2 is similar to, or slightly smaller than, that of NGC 3603.

\subsection{Age\label{age}}
Stars in young open clusters are considered to have the same age with a small age spread.
Therefore we can estimate the age of an open cluster from the morphology of the CMD or H-R diagram.
For a young open cluster, the MS turn-on is a good age indicator if the membership selection is reliable.
The evolutionary stages of the most massive stars and PMS stars have also been used in age estimations \citep{su97,hur12,li13}.

\begin{figure}
\begin{center}
\includegraphics[]{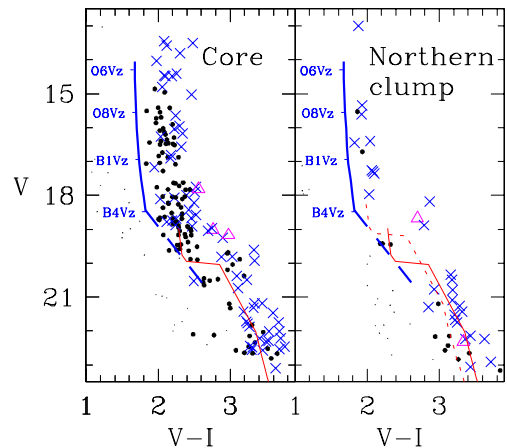}
\caption{The ($V,V-I$) CMDs of the stars with $V$-NIR colour in the core and in the northern clump.
The stars bluer than the cluster-field boundary or the reddened ZAMS of $E(B-V)=1.45$ in $V$-NIR colour are marked as small dots.
Crosses and triangles represent the X-ray emission stars and class II objects, respectively.
The thick solid and dashed lines represent the reddened ZAMS relation for $E(B-V)=1.45$ and $V_0=M_V=13.9$ mag, 
and the MS turn-on of $M_{V}=-0.6$ mag, respectively.
The thin solid and dashed lines represent the 1.5 Myr isochrones of the PMS evolutionary tracks of \citet{si00}
for the reddening of $E(B-V)=1.75$ and $E(B-V)=1.55$, respectively.}
\label{cmd_msto}
\end{center}
\end{figure}

Fig.~\ref{cmd_msto} shows the CMDs of the core and the northern clump. 
Most suspected field stars are well separated from cluster stars in Fig.~\ref{cmd_msto}.
The left panel of Fig.~\ref{cmd_msto} shows the ($V,V-I$) CMD of the most crowded region of Wd 2 (core) 
and therefore the region is expected to be least contaminated by field interlopers.
A clear MS turn-on at $M_{V}=-0.6$ mag ($\sim$B4 Vz, thick line) is seen from the stars in the core.
The MS turn-on is nicely matched to the 1.5 Myr isochrone from the PMS evolutionary tracks of \citet{si00}.
In the $(K_S,V-K_S)$ CMD, several PMS stars of class II stage are 2--3 mag brighter than the MS turn-on and appear to be Herbig Be stars.
The presence of such stars with strong excess in the $K_S$ band also support the very young MS turn-on age of 1.5 Myr.
\citet{va13} estimated an extremely young age of $\sim$0.6 Myr based on the size of the cavity.
A simple change in the adopted distance of 6 kpc gives an age of $\sim$1.2 Myr, which agrees with our age estimation.
The age of 1.5 Myr is also in good agreement with the previous age determinations of \citet{as07} ($2.0\pm 0.3$ Myr) and \citet{ca13} (1.5 Myr),
although their reddening and distance determinations are quite different from ours.

Some PMS stars in the northern clump are fainter than the 1.5 Myr isochrone shifted for the reddening of $E(B-V)=1.75$,
while the PMS stars in the core are brighter than the isochrone. 
The PMS stars in the northern clump are similarly fitted with the less reddened isochrone shifted by $E(B-V)=1.55$ 
which is similar to the reddening of MSP 18 ($E(B-V)=1.56$).
Therefore, the fainter PMS stars in the northern clump can be explained as having a smaller reddening rather than a younger age.
We therefore conclude that there is no clear evidence of an age difference between the core and the northern clump.

A comparison between the evolutionary stages of the most massive stars and the isochrone from an evolutionary model is another way to estimate the age of a cluster. 
However, as shown in the H-R diagram (Fig.~\ref{hrd}), there are no distinguishably evolved stars from the MS band. 
WR stars have often been considered as He-burning post-MS stars, but \citet{sm08} noted that some WNh stars could be the late-H-burning stages of very massive stars.
The 1.5 Myr isochrone from \citet{ek12} shows no significant difference from the ZAMS,
therefore it is not easy to estimate the age from massive stars only.
We simply confirm that the evolutionary stage of massive stars of Wd 2 also supports the age of 1.5 Myr estimated from the MS turn-on. 

Comparison between the present mass of WR 20a ($83\pm5$ and $82\pm5$ $M_{\sun}$ from \citealt{bo04})
and the most massive stellar evolutionary model ($120M_{\sun}$) of \citet{ek12} predicts an age of $\sim2.5$ Myr.
But we note that the light curve of WR 20a shows an evidence that it is a contact binary system (see Fig.5 of \citealt{ra07}).
If WR 20a had experienced mass transfer between its two components, this age estimation would be uncertain without consideration of such mass transfer.

There are two difficulties in an age estimation from PMS stars for Wd 2. 
The completeness estimations would have to reach down to low-mass (m $\lid 1 M_{\sun}$) stars 
in order to avoid the known overestimation of the age of intermediate-mass (1 -- 3 $M_{\sun}$) PMS stars \citep{su97,ha03}.
In Fig.~\ref{hrd}, the massive PMS stars seem to be younger than 1.5 Myr and some extremely bright stars seem to be younger than 0.3 Myr.  
But as noted by \citet{si00}, an age estimation using the PMS evolutionary tracks for ages younger than 1 Myr may be uncertain. 
Thus we just estimate the lower limit of the age of PMS stars in Wd 2 to be younger than 1 Myr.

\section{Discussion\label{disc}}
\subsection{The PMS stars in the halo of Westerlund 2\label{halo_pms}}
It is worth checking the age of the stars in the northern clump because it may be related to the star formation history within Wd 2.
But as we showed in Section~\ref{age}, we found no clear evidence of an age difference between the core and the northern clump,
therefore we conclude that the formation of both groups seems to be coeval.
There is a small filamentary structure connecting MSP 18 and a very red star ID 5744 (2MASS J10240187-5744152 -- G284.2600-00.3114),
but the star ID 5744 is too red ($V=20.58$ and $V-I=4.49$) to be a cluster member or an object related to the filament.

An unreported feature in the distribution of X-ray sources is the existence of more widely distributed sources around Wd 2.
Fig.1 of \citet{na08} shows the spatial distribution of X-ray point sources and 
clearly reveals the presence of distributed X-ray sources up to $\sim 2.5$ arcmin from the core.
Many of these distributed X-ray point sources seem to have moderate photon energies (1 -- 2 keV, green in Fig.1 of \citealt{na08})
and therefore they would seem to be PMS stars in the halo of Wd 2.
Several other compact young open clusters also seem to have such a halo. 
From a plateau in the surface density profile, \citet{ko10} identified a stellar halo associated with the compact young open cluster Hogg 15.
A very massive young open cluster, Westerlund 1, also shows a slight slope variation in its radial profile and a radial variation in the IMF \citep{li13}.
Such a radial variation of the IMF is also seen in the massive young open cluster, NGC 3603, which is very similar in mass and age \citep{sb04,st06} to Wd 2.
Wd 2 also shows a difference in the slope of the IMF between the core ($\Gamma=-1.0\pm0.1$) and the halo ($\Gamma=-1.5\pm0.2$).
Therefore many of the X-ray sources in the halo of Wd 2 seem to be low-mass PMS stars that are widely distributed as a result of mass segregation processes. 

\begin{figure*}
\includegraphics[height=0.47\textwidth]{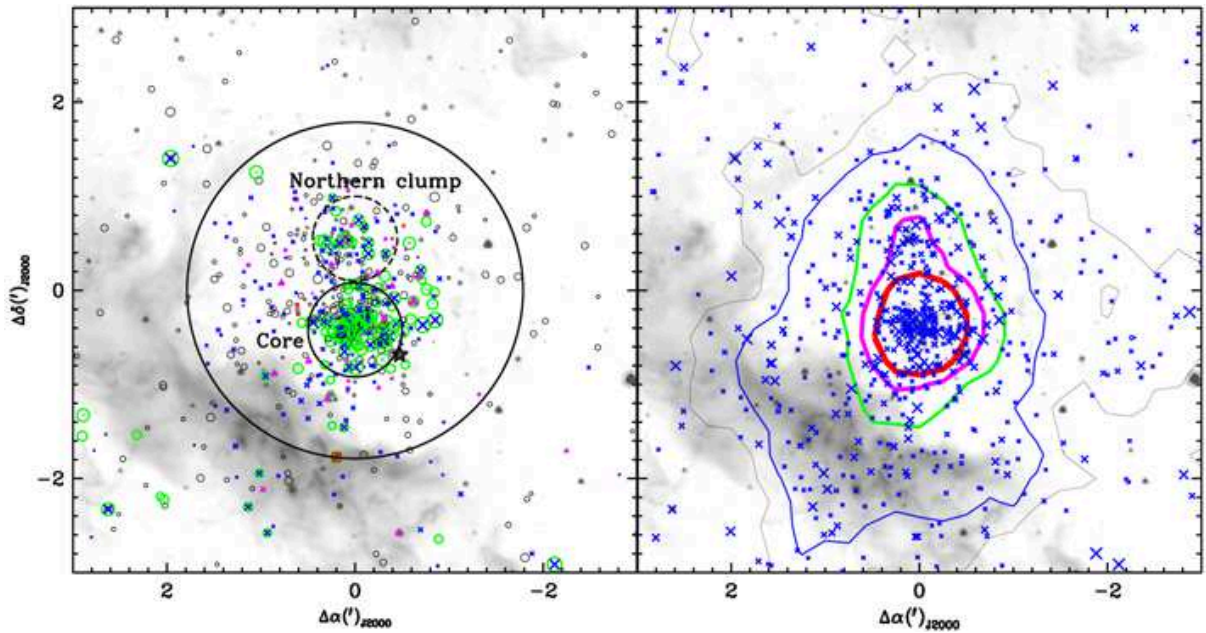}
\caption{
Spatial distribution of members and X-ray point sources on the reversed {\it Spitzer} 4.5$\mu m$ image in which the bright bridge is well highlighted.
{\it left} -- The distribution of members and candidates around Wd 2.
The black circles are the PMS candidates and the other symbols are the same as Fig.~\ref{redmap}.
{\it right} -- The distribution of {\it Chandra} X-ray sources. The size of symbols are based on $\log(significance)$.
The thin to thick contours indicate number densities of X-ray sources of 7, 15, 40, 70, and 100 per an arcmin$^2$, respectively.}
\label{halo_cl}
\end{figure*}

Another remarkable feature is that the X-ray sources in the halo seem to be more distributed toward the south-east of the core.
In the right panel of Fig.~\ref{halo_cl}, 
the location of the distributed X-ray sources at the south-east of the core are fully overlapped with the bright bridge of the RCW 49 nebula 
[($\Delta \alpha \sim 2\farcm0$, $\Delta \delta \sim -0\farcm5$) -- ($\Delta \alpha \sim -1\farcm0$, $\Delta \delta \sim -2\farcm0$)].
Although an extended distribution of faint stars toward the bridge is not clearly seen in the optical data
due to the relatively higher reddening at the southern edge of Wd 2,
a number of the early-type stars are also found at the edge of the bridge.
As only a few of these X-ray sources in the bridge were identified with optical sources,
it is not easy to analyze the properties of these X-ray sources using the optical data.
The bridge spatially coincides with the most redshifted cloud \citep{fu09} 
which implies the bridge is not a foreground cloud to Wd 2
and is not the main reason for the lower detection of the X-ray sources in the optical.
Strong Br$\alpha$ emission \citep{ch05} and PAH emission (Fig. 10 of \citealt{be13}) in the bridge toward Wd 2 (north-west edge of the bridge)
are also confirmed.
\citet{be13} suggested the presence of a non-thermal radio emission in the bridge.
A possible scenario is that these X-ray sources may be the PMS stars recently formed as a consequence of sequential star formation 
triggered by the strong radiation field of Wd 2 and that they are highly obscured by circumstellar dust or material remaining in the bridge.
Although X-ray emission is a good tracer of PMS stars at the class II stage, many of younger PMS stars, such as class I stars, also show X-ray emission.
For example, in the case of the young open cluster NGC 2264 \citep{su09}, 
41 out of 375 (11\%) class II stars were identified as X-ray point sources while 19 out of 89 (21\%) class I stars were identified as X-ray point sources.
This scenario also supports the possibility of the bright bridge being a region compressed on both sides by Wd 2 and WR 20b,
as suggested by \citet{wu97} and \citet{ch04}.

\subsection{The Distribution of Early-type Stars in RCW49 ($r>r_{cl}$) -- An OB Association? \label{halo_early}}
In contrast to the distribution of X-ray point sources, 
the early-type stars outside $r_{cl}$ are distributed farther from Wd 2 and seem to be linearly aligned 
from ($\alpha_{J2000}\sim 10.411$, $\delta_{J2000}\sim -57.755$)
to ($\alpha_{J2000}\sim 10.392$, $\delta_{J2000}\sim -57.85$) (Fig.~\ref{redmap}).
This alignment is well matched to the large filamentary structure extended from the east to the south-west of Wd 2 in the MIR images.
Several of the early-type stars outside $r_{cl}$ could be divided into small groups.
A small group near ($\alpha_{J2000}\sim 10.411$, $\delta_{J2000}\sim -57.755$) contains six early-type stars.
Two of these stars, ID 8885 and 8904, are confirmed to be late-O type stars from our spectroscopic observations
and an early-O star ID 8885 ($<$O6) shows a bow shock toward Wd 2 (object S3 in \citealt{po08}).
There is a very compact early-type stellar group near ($\alpha_{J2000}\sim 10.392$, $\delta_{J2000}\sim -57.85$) with five early-type stars 
(ID 2914, 2951, 2954, 2972, and 3000).
The star ID 2972 is confirmed to be a late-O star from our spectroscopic observations and other stars seem to be early-B stars based on their reddening-corrected colours.
These two groups may be isolated groups of stars in the X-ray image (see Fig. 1 of \citealt{na08}).

We can consider two possibilities for the origin of the early-type stars outside $r_{cl}$, 
(1) the dynamical ejection due to the encounters among stars in the core of Wd 2, and (2) in situ star formation.
Recent N-body simulations \citep{ok12,fz12} predict that the ejection of massive stars is possible 
by dynamical encounters such as star-star, binary-star, or binary-binary interactions in a massive crowded cluster.
Recently \citet{ro11} reported that WR 20aa (O2 If*/WN6), WR 20c (O2 If*/WN6), and WR 21a (O3 If*/WN6 + early-O) are possibly ejected massive stars from Wd 2 
based on the geometrical alignment between these stars and Wd 2.
The results of these studies support the possibility of the presence of early-type stars outside $r_{cl}$
which were initially formed within Wd 2 and ejected by dynamical encounters.
But the alignment of the early-type stars along the bright filamentary structure on the MIR images and 
the presence of the compact early-type stellar groups cannot be fully explained by the dynamical encounter scenario.
Furthermore, the early-type stars outside $r_{cl}$ are located in the cavities of the RCW 49 nebula in the MIR images.
Therefore, many of the early-type stars outside $r_{cl}$, rather than being ejected by dynamical encounters, seem to be formed in situ  
and members of an OB association in the RCW 49 nebula.

Although we selected several OB stars outside $r_{cl}$ from the CMDs and ($U-B,B-V$) TCD, the presence of additional embedded OB stars is also possible.
For example, RCW 49-S2 in \citet{po08} [ID 5911 ($V=19.72$, $B-V=2.39$ $V-I=3.62$) = 2MASS J10240306-5748356] 
shows a bow shock and seems to be an O6 -- O9 V star based on their SED fitting.
RCW 49-S3 may also be an O-type star, as there is an agreement between their spectral type (O3 -- O5 V) based on SED fitting 
and our spectroscopic observation (ID 8885, $<$O6). However, RCW 49-S2 was not selected as an early-type star in our analysis
because we had to discard its $U-B$ due to the uncorrectable red leak effect for $B-V>1.7$ ($E(B-V)\ga 2.0$).
If we assume the star is an early-type star, a de-reddening on the ($V,B-V$) CMD yields $E(B-V)=2.72$ and $(B-V)_0=-0.33$ which implies it is an early-O (O2 -- O4) star
while the spectral type from \citet{po08} (O6 -- O9 V) implies $E(B-V)=2.6-2.7$. 
In either case, the star is most likely a highly reddened O-type star which was not selected from our optical photometric data.
Therefore, we expect the existence of more embedded OB stars in the RCW 49 nebula.

\section{Summary\label{summ}}
For stars in the young starburst-type cluster Wd 2 and the RCW 49 nebula, we have presented deep $UBVI_C$ photometric data down to $V=23$ mag 
obtained with the the MOSAIC II CCD camera on the 4m Blanco telescope at CTIO.
Optical photometric data as well as $JHK_S$ data from \citet{as07} and the 2MASS point source catalog were used for the membership selection.
We also used the {\it Chandra} X-ray point source list from \citet{na08} and the GLIMPSE point source catalog for the selection of PMS members.

We confirmed a reddening dependent term of $E(U-B)/E(B-V)$ ratio ($=0.72+0.025E(B-V)$) from the spectral type--intrinsic colour relation, and  
determined an abnormal reddening law of $R_{V,cl}$ = $4.14\pm0.08$ for the early-type stars in Wd 2
from $E(V-J)/E(B-V)$, $E(V-H)/E(B-V)$, and $E(V-K_S)/E(B-V)$ ratios.
For the foreground stars with $E(B-V)_{fg}<1.05$, a fairly normal reddening law $R_{V,fg}=3.33\pm0.03$ was also confirmed.
The distance modulus was determined to be $V_0-M_V=13.9$ mag $\pm$0.1 (from photometric error) $\pm$0.1 (from reddening law error)
$_{-0.00}^{+0.40}$ (the systematic scatter of the ZAMS) (d=6.0 kpc $\pm$0.3 $\pm$0.3 $_{-0.3}^{+1.2}$) from ZAMS fitting to the reddening-corrected CMD.

The age of Wd 2 was estimated to be $\la1.5$ Myr from the MS turn-on and the isochrones from the PMS evolutionary tracks. 
The IMF of Wd 2 was derived down to $\log m = 0.7$ and the global slope of the IMF was obtained to be $\Gamma=-1.1\pm0.1$.
The lower limit of the total mass of Wd 2 is about $7,400 M_{\sun}$, and the cluster seems to be as massive as the young galactic starburst-type cluster NGC 3603.

Many faint PMS stars were detected in and around the MIR ring near the massive star MSP 18 located to the north of the core,
but we could not find clear evidence for an age difference between the core and the northern clump.
We report that Wd 2 has an extended halo, with a radius of more than 2 arcmin, composed mainly of low-mass PMS stars.
The halo is more extended toward the south-east and fully overlapped with the bright bridge of the RCW 49 nebula.
We suggest that the stars in the extended halo may be newly formed PMS stars as a result of sequential star formation. 
Finally, we discovered several early-type stars in the RCW 49 nebula.
These early-type stars are well aligned along the bright filamentary structure in the MIR image of the RCW 49 nebula and seem to be formed in situ.
These early-type stars appear to be members of an OB association in the RCW49 nebula.

\section*{Acknowledgements}
This work is partly supported by a National Research Foundation of Korea (NRF) grant funded by the Korean Government (MEST) (Grant No. 2013031015)
and partly supported by the Korea Astronomy and Space Science Institute (KASI) (Grant No. 2011-9-300-02).

\appendix
\section{A Non-linear Term in the Transformation of the $U$ Filter of the CTIO 1m Telescope\label{f_U_text}}
Compared with the other optical pass bands, standardization of $U$ and $U-B$ is more difficult for several reasons.
One of the difficulties is a non-linear correction related to the Balmer discontinuity 
which depends on the intrinsic colour (or spectral type) of the stars (see \citealt{bs90} and references therein).
The non-linear correction term has been frequently confirmed in the transformations of data obtained from many instruments \citep{su00,li09}.
Therefore, in the standardization of photometric data with filters which transmit over the 3700\AA -- 3800\AA $ $ wavelength range, 
the presence of a non-linear correction term to reproduce the standard colour should be tested for.

\begin{figure}
\begin{center}
\includegraphics[height=0.2\textwidth]{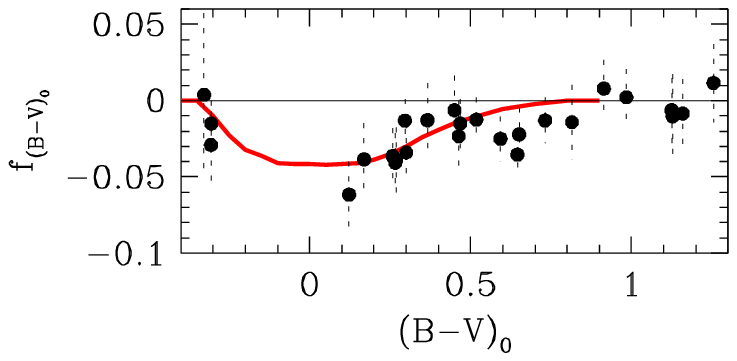}
\caption{The residual after the standardization 
($U - u +[k_{1U}-k_{2U}(U-B)]X-\eta _U(U-B)-\alpha _UUT-\gamma_{1U}r-\gamma_{2U}r^2-\zeta_U$) of the SAAO standard stars observed with the CTIO 1m telescope.
The dashed lines are the standard deviation of the observed standard stars.
The thick solid line (red) indicates the non-linear correction term ($f_{[(B-V)_0]}$) in the transformation of the $U$ filter of the CTIO 1m telescope.}
\label{f_1mU}
\end{center}
\end{figure}

We tested for the presence of a non-linear correction term for the CTIO 1m $U$ filter using the SAAO E-region standard stars \citep{me89,ki98}
and a small ($\lid 0.042$ mag) correction term as a function of the intrinsic $B-V$ colour was confirmed.
Fig.~\ref{f_1mU} shows the non-linear correction term ($f_{[(B-V)_0]}$) of the CTIO $U$ filter transformation.
The peak of the non-linear term is at $(B-V)_0=0$ (A0).

\section{Transmission of the $B$ Filter of the CTIO 1m Telescope}
\begin{figure}
\begin{center}
\includegraphics[height=0.50\textwidth]{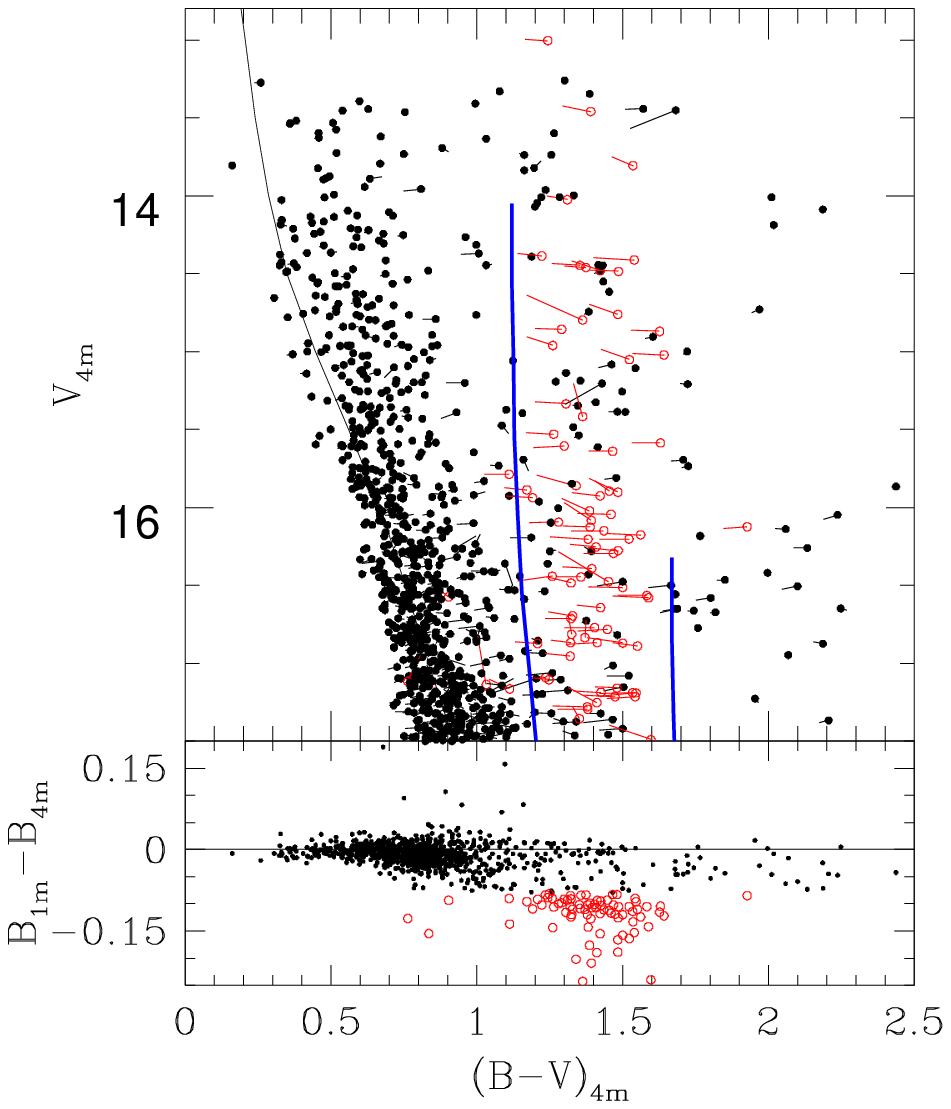}
\caption{The CMD and the difference in the $B$ magnitudes between the CTIO 1m and 4m for $V\protect\lid 17.5$ mag with a photometric error of
$\epsilon(\equiv \sqrt{{\epsilon V}^2+{\epsilon (B-V)}^2}){\protect\lid} 0.1$.
Photometric doubles were excluded here. 
The reddened ZAMS appropriate for the foreground field stars [thin line: $E(B-V)=0.15$ \& $V_0-M_V=11.3$] 
and for Wd 2 [thick lines: $E(B-V)=1.45$ \& $V_0-M_V=13.9$, and $E(B-V)=2.00$ \& $V_0-M_V=13.9$] are shown in the figure.
Open circles are the stars with $B_{1m}-B_{4m} < -0.08$.
Short lines in the ($V,B-V$) CMD (upper panel) represent the differences of position in the CMD between the 1m and 4m photometry.}
\label{ctio1b_cmd}
\end{center}
\end{figure}

\begin{figure}
\begin{center}
\includegraphics[height=0.20\textwidth]{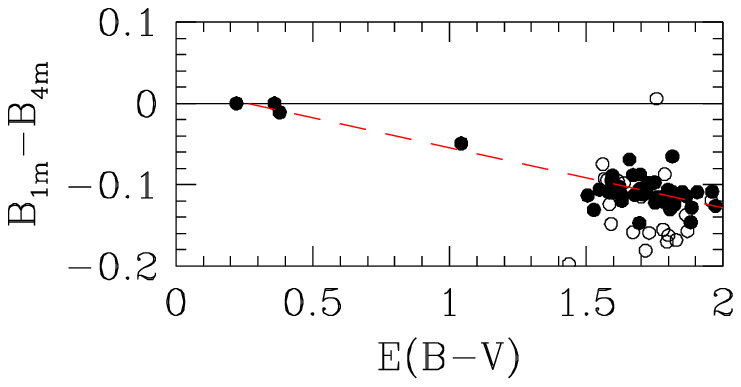}
\caption{The difference in the $B$ magnitude of the early-type stars. 
Open circles represent photometric doubles, stars possibly affected by the red leak of a close companion,
or variable candidates ($\Delta V =|V_{1m}-V_{4m}| {\protect\gid} 0.03$). 
The red dashed line represents the relation between the two $B$ magnitudes: $B_{1m}-B_{4m}= -0.075 \times E(B-V)+0.020$}
\label{ctio1b_ebv}
\end{center}
\end{figure}

\begin{figure}
\begin{center}
\includegraphics[height=0.35\textwidth]{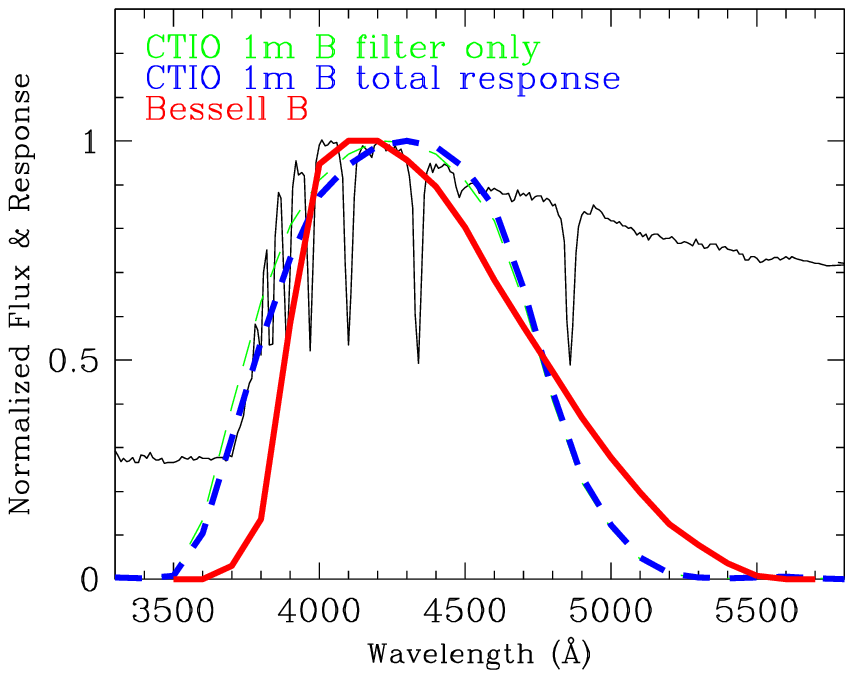}
\caption{Comparison of the total response of the CTIO 1m $B$ band (thick dashed line) and the standard $B$ band (thick solid line, \citealt{bs12}).
The thin dashed line represent the transmission of the CTIO 1m $B$ filter only.
A spectral energy distribution of an A0 V star provided by \citet{gs83} is shown for the comparison (thin solid line).}
\label{ctio1b_tr}
\end{center}
\end{figure}

Although the $B$ magnitudes of the stars in our field of view showed good agreements between CTIO 1m and 4m data,
highly reddened stars showed a significant systematic difference in the $B$ magnitude.
In Fig.~\ref{ctio1b_cmd}, the sequences of stars in the Sagittarius--Carina arm ($E(B-V)\sim 0.3$, black thin line in Fig.~\ref{ctio1b_cmd})
show a consistency between the two $B$ band observations.
However, stars in the sequence of Wd 2 ($1.45\lid E(B-V) \lid 2.00$, within blue thick lines) show large differences in the $B$ band.
As there are two clear sequences in the lower panel of Fig.~\ref{ctio1b_cmd}, it seems to be not a simple colour dependence.
This unusual difference in the $B$ magnitudes appears to be caused by a difference in reddening (see Fig.~\ref{ctio1b_ebv}).
It is not easy to explain why the difference has arisen, but a possible cause is the abnormally broad transmission of the CTIO 1m $B$ filter.
Fig.~\ref{ctio1b_tr} shows a comparison of the total response of the CTIO 1m $B$ band\footnote{http://www.astronomy.ohio-state.edu/Y4KCam/filters.html}
and the standard $B$ band \citep{bs12}.
For the calculation of the total response of the CTIO 1m $B$ band, 
we assumed the typical reflectivity of an aluminum coating (nearly constant within the transmission of $B$ filter) for the primary and the secondary mirrors and 
adopted the quantum efficiency of the Y4KCAM\footnote{http://www.astronomy.ohio-state.edu/Y4KCam/OSU4K/index.html}.
The two $B$ bands show a significant difference in response, especially in the wings.
Although it is not easy to explain clearly which parameter causes the difference in the $B$ magnitudes of highly reddened stars,
the difference does depend on reddening.
But it is certain that the redder wing of the CTIO 1m $B$ filter and the ultraviolet cut off that is unusually extended toward short wavelength
seems to be affected by the Balmer jump and the confluence of the Balmer lines,
while its central wavelength is shifted toward longer wavelength.
Therefore a rigorous check of data obtained with the CTIO 1m $B$ filter seems to be necessary.

\end{document}